\documentclass[aps,prl,twocolumn,superscriptaddress]{revtex4-1}

\makeatletter
\def\@fnsymbol#1{\ensuremath{\ifcase#1\or \dagger\or *\or \ddagger\or
   \mathsection\or \mathparagraph\or \|\or **\or \dagger\dagger
   \or \ddagger\ddagger \else\@ctrerr\fi}}
\makeatother

\usepackage{dcolumn}
\usepackage{amsfonts}
\usepackage{amsmath}
\usepackage{amssymb}
\usepackage{amsthm}
\usepackage{mathrsfs}
\usepackage{amsbsy}
\usepackage{amstext}
\usepackage{siunitx}
\usepackage{fixltx2e}
\usepackage{upgreek,xspace}
\usepackage{multirow}
\usepackage{wasysym}
\usepackage{mathtools}
\usepackage{bbold,bm}
\usepackage{graphicx}
\usepackage[colorlinks=true,citecolor=blue,linkcolor=blue,urlcolor=blue]{hyperref}
\usepackage{bm}
\usepackage{color}
\usepackage[T1]{fontenc}

\newcommand{\beginextdata}{%
        \setcounter{figure}{0}
        \renewcommand{\thefigure}{Extended Data~\arabic{figure}}%
     }

\newcommand{\beginsupplement}{%
        \setcounter{table}{0}
        \renewcommand{\thetable}{S\arabic{table}}%
        \setcounter{figure}{0}
        \renewcommand{\thefigure}{S\arabic{figure}}%
        \setcounter{equation}{0}
        \renewcommand{\theequation}{S\arabic{equation}}%
     }

\begin{document}

\title{\Large Discovery of the soft electronic modes of the trimeron order\\ in magnetite}

\author{Edoardo Baldini}
\altaffiliation{These authors contributed equally to this work.}
\affiliation{Department of Physics, Massachusetts Institute of Technology, Cambridge, Massachusetts 02139, USA}

\author{Carina A. Belvin}
\altaffiliation{These authors contributed equally to this work.}
\affiliation{Department of Physics, Massachusetts Institute of Technology, Cambridge, Massachusetts 02139, USA}
	
\author{Martin Rodriguez-Vega}
\affiliation{Department of Physics, The University of Texas at Austin, Austin, Texas 78712, USA}
\affiliation{Department of Physics, Northeastern University, Boston, Massachusetts 02115, USA}

\author{Ilkem Ozge Ozel}
\affiliation{Department of Physics, Massachusetts Institute of Technology, Cambridge, Massachusetts 02139, USA}
	
\author{Dominik Legut}
\affiliation{IT4Innovations Center, VSB-Technical University of Ostrava, 17.listopadu 15, 708 00 Ostrava, Czech Republic}
	
\author{Andrzej Koz\l{}owski}
\affiliation{Faculty of Physics and Applied Computer Science, AGH-University of Science and Technology,\\ Aleja Mickiewicza 30, PL-30059 Krak\'{o}w, Poland}
	
\author{Andrzej M. Ole\'{s}}
\affiliation{Marian Smoluchowski Institute of Physics, Jagiellonian University, Prof. S. \L{}ojasiewicza 11, PL-30348 Krak\'{o}w, Poland}
\affiliation{Max Planck Institute for Solid State Research, Heisenbergstra{\ss}e 1, D-70569 Stuttgart, Germany}

\author{Krzysztof Parlinski}
\affiliation{Institute of Nuclear Physics, Polish Academy of Sciences, Radzikowskiego 152, PL-31342 Krak\'{o}w, Poland}
	
\author{Przemys\l{}aw Piekarz}
\affiliation{Institute of Nuclear Physics, Polish Academy of Sciences, Radzikowskiego 152, PL-31342 Krak\'{o}w, Poland}
	
\author{Jos\'e Lorenzana}
\affiliation{Institute for Complex Systems, National Research Council and Department of Physics, University of Rome ``La Sapienza," I-00185 Rome, Italy}
	
\author{Gregory A. Fiete}
\affiliation{Department of Physics, Massachusetts Institute of Technology, Cambridge, Massachusetts 02139, USA}
\affiliation{Department of Physics, Northeastern University, Boston, Massachusetts 02115, USA}

\author{Nuh Gedik}
\altaffiliation{Email: gedik@mit.edu}
\affiliation{Department of Physics, Massachusetts Institute of Technology, Cambridge, Massachusetts 02139, USA}
	
\date{February 2, 2020}

\maketitle

\noindent\textbf{The Verwey transition in magnetite (Fe$_3$O$_4$) is the first metal-insulator transition ever observed \cite{verwey1939electronic} and involves a concomitant structural rearrangement and charge-orbital ordering. Due to the complex interplay of these intertwined degrees of freedom, a complete characterization of the low-temperature phase of magnetite and the mechanism driving the transition have long remained elusive. It was demonstrated in recent years that the fundamental building blocks of the charge-ordered structure are three-site small polarons called trimerons \cite{senn2012charge}. However, electronic collective modes of this trimeron order have not been detected to date, and thus an understanding of the dynamics of the Verwey transition from an electronic point of view is still lacking. Here, we discover spectroscopic signatures of the low-energy electronic excitations of the trimeron network using terahertz light. By driving these modes coherently with an ultrashort laser pulse, we reveal their critical softening and hence demonstrate their direct involvement in the Verwey transition. These findings represent the first observation of soft modes in magnetite and shed new light on the cooperative mechanism at the origin of its exotic ground state.}

Along with his groundbreaking discovery in 1939, Verwey postulated the emergence of a charge ordering of Fe$^{2+}$ and Fe$^{3+}$ ions as the mechanism driving the dramatic conductivity drop at $T_V \sim$ 125 K \cite{verwey1939electronic}. A vast number of subsequent experimental and theoretical investigations, including those by Anderson \cite{anderson1956ordering}, Mott \cite{mott1980materials}, and many others, have stimulated a still unresolved debate over a complete description of the Verwey transition \cite{walz2002verwey,khomskii2014transition}. In particular, several seemingly incompatible findings related to the intricate low-temperature phase of magnetite have been reported: the crucial role of Coulomb repulsion \cite{leonov2004charge}, the necessity of including electron-phonon coupling \cite{mott1980materials,yamada1980molecular,piekarz2006mechanism}, small charge disproportionation \cite{wright2001long, leonov2004charge,subias2012structural}, anomalous phonon broadening with the absence of a softening towards $T_V$ \cite{hoesch2013anharmonicity}, and the observation of structural fluctuations that are connected to the Fermi surface nesting \cite{bosak2014short} and that persist up to the Curie transition temperature ($T_C$ $\sim$ 850 K) \cite{perversi2019co}.

\begin{figure}[htb!]
\includegraphics[width=\columnwidth]{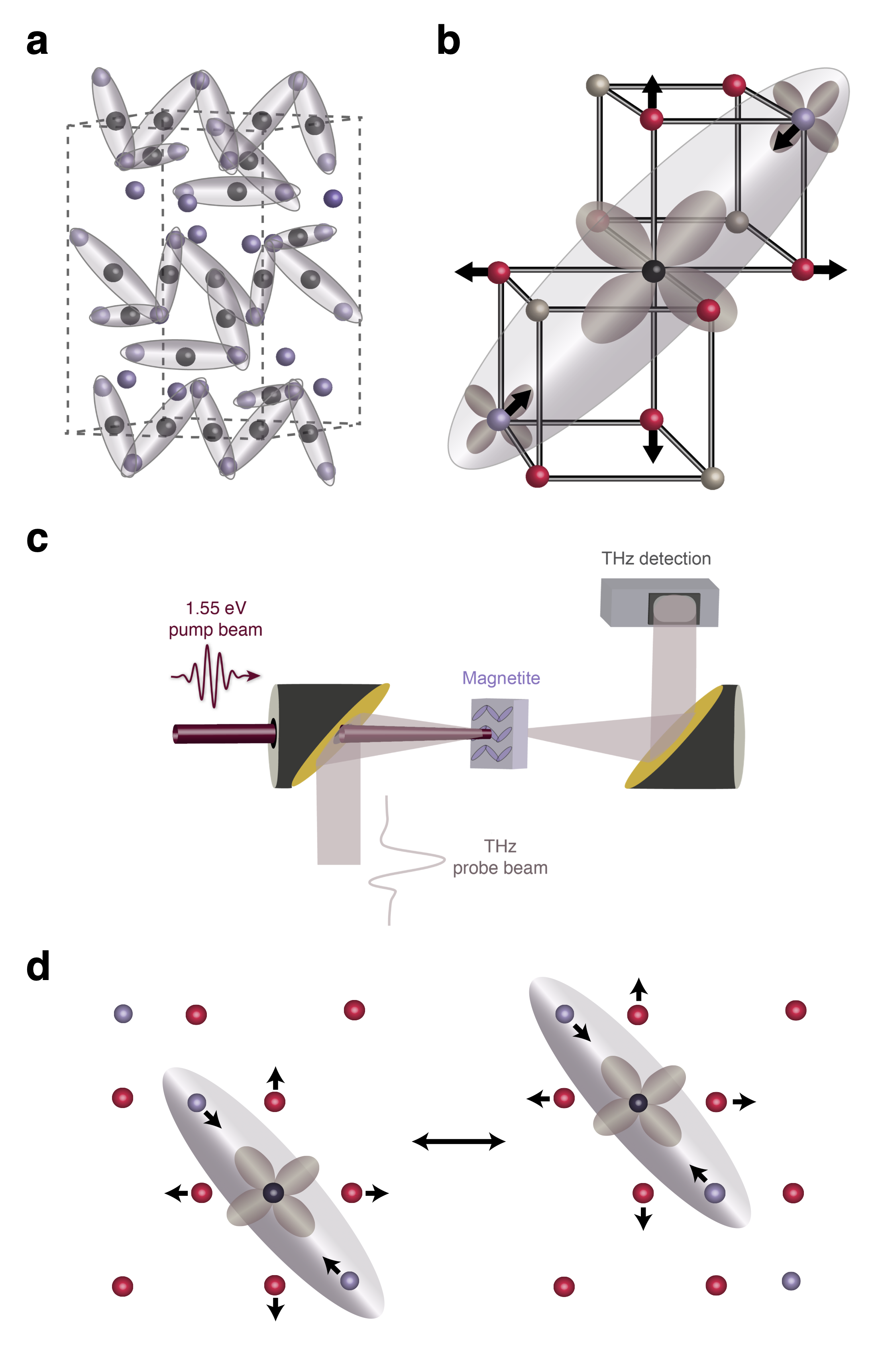}
\caption{\textbf{Trimeron order in magnetite and experimental methodology.} \textbf{a}, The low-temperature charge-ordered structure of magnetite as a network of trimerons, small polarons that extend over three linear Fe sites. The purple and black spheres represent Fe$^{3+}$ and Fe$^{2+}$ ions, respectively. \textbf{b}, Each trimeron consists of two outer Fe$^{3+}$ ions and one central Fe$^{2+}$ ion (the arrows depict distortions of the lattice). This charge ordering is accompanied by an ordering of coplanar $t_{2g}$ orbitals on each Fe site within the trimeron. The surrounding oxygen ions are shown in red, and the gray spheres represent Fe sites that do not participate in the trimeron. (\textbf{a} and \textbf{b} are adapted from Ref. \cite{senn2012charge}.) \textbf{c}, Schematic of the experimental setup. Time-domain THz spectroscopy in a transmission geometry is used to determine the low-energy optical conductivity of the sample in equilibrium (without the pump beam). To examine its dynamics, an ultrashort near-infrared (1.55 eV) pump pulse sets the system out of equilibrium and a weak, delayed THz probe pulse measures the pump-induced change in the optical conductivity. \textbf{d}, Cartoon depicting the trimeron sliding motion resulting from the coherent polaron tunneling between an Fe$^{2+}$ site and a neighboring Fe$^{3+}$ site.}
\label{fig:Fig1}
\end{figure}

The last decade witnessed significant progress in understanding the Verwey transition from a structural point of view. Most notably, a refinement of the low-temperature charge-ordered structure as a network of three-site small polarons, termed trimerons, was given by x-ray diffraction \cite{senn2012charge} (Fig.~\ref{fig:Fig1}a). A trimeron consists of a linear unit of three Fe sites accompanied by distortions of the two outer Fe$^{3+}$ ions towards the central Fe$^{2+}$ ion. An orbital ordering of coplanar $t_{2g}$ orbitals is also established on each ion within the trimeron (Fig.~\ref{fig:Fig1}b). This picture of the trimeron order has been crucial for determining the correct noncentrosymmetric $Cc$ space group of magnetite and explaining its spontaneous charge-driven ferroelectric polarization \cite{yamauchi2009ferroelectricity,senn2012charge,khomskii2014transition}. Nevertheless, despite extensive research, no soft modes of the trimeron order have been detected to date. Unveiling novel types of collective modes in the low-temperature phase of magnetite and their critical softening would significantly shape our understanding of the long-sought cooperative phenomenon at the origin of the Verwey transition.

Here, we use time-domain terahertz (THz) spectroscopy (Fig.~\ref{fig:Fig1}c) to reveal the electronic modes of the trimeron order. Their signature is imprinted on the equilibrium optical conductivity of the material in an energy-temperature range previously unexplored. We establish their involvement in the Verwey transition by driving them coherently with an ultrashort near-infrared laser pulse and mapping their softening with a delayed THz probe. We propose a model of coherent polaron tunneling to describe the nature of these trimeron excitations.

\begin{figure*}[t!]
\includegraphics[width=\textwidth]{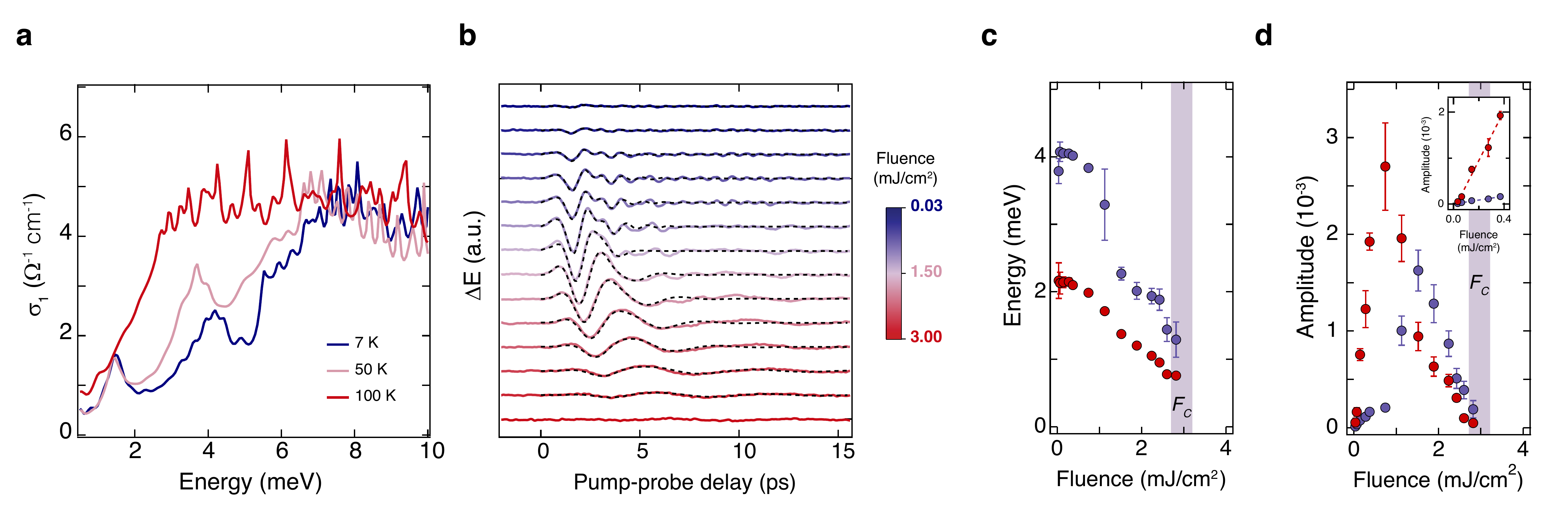}
\caption{\textbf{Observation of low-energy electronic collective modes and their critical softening.} \textbf{a}, Real part of the equilibrium optical conductivity ($\sigma_1$) of magnetite in the THz region. The spectrum near $T_V$ (red curve) exhibits a broad, featureless continuum that rises with a power-law behavior, in accordance with a previous report \cite{pimenov2005terahertz}. Lowering the temperature reveals two Lorentzian lineshapes (pink and blue curves) that are due to collective modes. \textbf{b}, Pump-induced change in the THz electric field ($\Delta E$) transmitted through the sample at 7 K following photoexcitation with a near-infrared (1.55 eV) pump pulse at various absorbed fluences. The traces are offset vertically for clarity. Each curve was fit to two damped sine waves and the fits in the time domain are displayed as dashed black lines. \textbf{c}, Energy of each oscillation extracted from the fits as a function of fluence. At low fluence, the energies are close to those in \textbf{a} at 7 K, so the oscillations correspond to the same collective modes present in equilibrium. Both energies soften with increasing fluence, demonstrating their involvement in the Verwey transition. \textbf{d}, Amplitude of each mode versus fluence. The inset shows the linear rise of the amplitudes in the low fluence regime, which is compatible with an impulsive Raman excitation process. The shaded bars in \textbf{c} and \textbf{d} indicate the critical fluence for melting the trimeron order as reported in Refs. \cite{de2013speed,randi2016phase}. The error bars in \textbf{c} and \textbf{d} represent the 95\% confidence interval for the corresponding fit parameters.}
\label{fig:Fig2}
\end{figure*}

Figure~\ref{fig:Fig2}a shows the real part of the low-energy optical conductivity ($\sigma_1$) measured in equilibrium on a magnetite single crystal. Slightly below $T_V$, the spectrum displays a broad, featureless continuum (red curve), which previous studies attributed to a power-law behavior expected in the presence of charge hopping between polaronic states \cite{pimenov2005terahertz}. As the temperature is lowered well below $T_V$ to a hitherto unexplored regime (pink and blue curves), the continuum in the optical conductivity is suppressed and two Lorentzian lineshapes clearly emerge. These excitations slightly harden with decreasing temperature and are centered around 1.5 and 4.2 meV at 7 K. Since they appear below the charge gap for single-particle excitations \cite{gasparov2000infrared}, it is natural to ascribe them to distinct low-energy collective modes at the Brillouin zone center. Intriguingly, these excitations have never been observed in any previous study on magnetite \cite{pimenov2005terahertz,mcqueeney2005influence,borroni2017mapping,huang2017jahn,borroni2018light,elnaggar2018site}. Thus, it is pivotal to identify their origin and clarify their potential involvement in the Verwey transition.

We make use of advanced density functional theory (DFT) calculations of the phonon dispersions in the low-temperature $Cc$ symmetry of magnetite to compare the energy of the two observed excitations with that of long-wavelength lattice modes (see Methods and Fig.~\ref{fig:FigEx1}a). The lowest-lying optical phonons at the $\Gamma$ point of the Brillouin zone have symmetries $A^\prime$ and $A^{\prime\prime}$ and correspond to the folded $\Delta_5$ mode of the cubic phase. As shown in Fig.~\ref{fig:FigEx1}a, their energy of 8 meV is in excellent agreement with inelastic neutron \cite{samuelsen1974low,borroni2017mapping} and x-ray \cite{hoesch2013anharmonicity} scattering data. Therefore, the low-energy modes in our experiment cannot be assigned to phonons. Similarly, magnon dispersions measured by inelastic neutron scattering do not show any long-wavelength spin waves with energies in our spectral range \cite{mcqueeney2005influence}. Since ferrimagnetism in magnetite is quite robust (with $T_C \sim$ 850 K), these excitations would be expected to persist at high temperature \cite{pimenov2005terahertz}. Thus, after ruling out these scenarios of conventional types of collective modes (see Supplementary Note 1 for additional details), the only remaining possibility is that the detected modes are collective excitations of the trimeron order.

We now clarify whether these collective modes play a key role in the Verwey transition by unraveling their critical behavior. This is a challenging task to accomplish under equilibrium conditions, as the broad conductivity continuum seen in Fig.~\ref{fig:Fig2}a obscures any spectroscopic signature of the collective modes at temperatures proximate to $T_V$. It is thus unclear whether the modes persist as strongly damped Lorentzians buried under this continuum and how their peak energies change with temperature. To overcome this experimental difficulty, we illuminate our magnetite crystal with an ultrashort near-infrared laser pulse and drive its collective modes coherently \cite{stevens2002coherent}. We vary the laser fluence absorbed by our sample, exploring a regime that allows us to transiently increase the lattice temperature but not completely melt the trimeron order \cite{de2013speed,randi2016phase} (see Supplementary Note 4). We then monitor the fingerprint of the coherently excited collective modes on the low-energy electrodynamics of the system.

\begin{figure*}[t!]
\includegraphics[width=\textwidth]{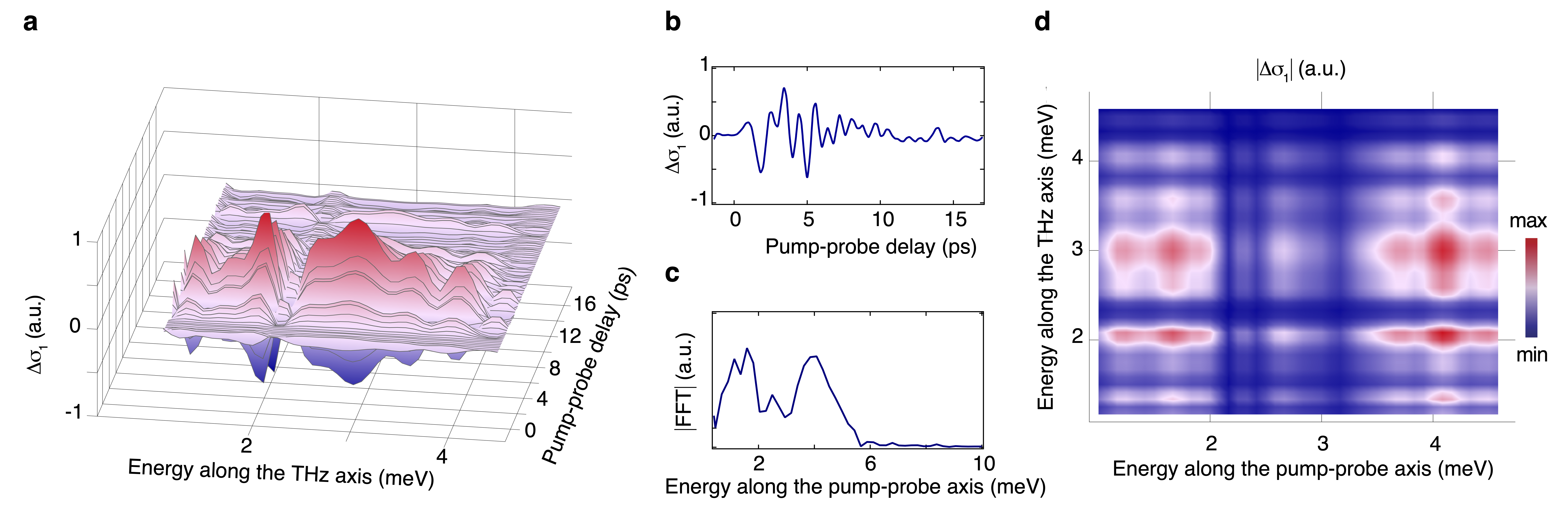}
\caption{\textbf{Mutual coupling between the two collective modes.} \textbf{a}, Spectro-temporal evolution of the photoinduced change in the real part of the optical conductivity ($\Delta\sigma_1$) in the THz range upon photoexcitation with a near-infrared (1.55 eV) pump pulse. The temperature is 7 K and the absorbed fluence is 0.27 mJ/cm$^2$. Two broad features are observed around 2 and 3 meV and they are coherently modulated as a function of pump-probe delay. \textbf{b}, Trace of $\Delta\sigma_1$ as a function of pump-probe delay at a representative THz photon energy. \textbf{c} Fourier transform of the temporal trace shown in \textbf{b}, which reveals two distinct peaks corresponding to the two coherent collective modes (compare their energies with Fig.~\ref{fig:Fig2}a,c). \textbf{d}, Amplitude of the Fourier transform taken along the pump-probe delay axis at each THz photon energy of the map shown in \textbf{a}. We observe that two peaks are present in the Fourier transform at each photon energy, which indicates that the two electronic modes are coupled to each other.}
\label{fig:Fig3}
\end{figure*}

Figure~\ref{fig:Fig2}b shows the pump-induced change in the THz electric field ($\Delta E$) through the sample over a range of absorbed fluences. As a function of pump-probe delay, we observe prominent oscillations that are signatures of collective modes coherently evolving in time. Upon closer examination, these oscillations cannot be described by a single frequency, but rather by two frequencies, which indicates the presence of two distinct coherent excitations (see Supplementary Note 5B). Fitting the time domain traces to the simplest model (consisting of two damped sine waves) reveals that at the lowest fluence the modes have energies very close to those present in the equilibrium conductivity at 7 K and therefore are the same excitations. Slight deviations in their energies relative to equilibrium possibly occur due to the non-thermal action of the pump pulse during photoexcitation. Their simultaneous presence in the frequency domain (as dipole-allowed modes) and in the time domain (as Raman-active modes) is naturally explained by the breakdown of the selection rules for infrared and Raman activity in the noncentrosymmetric $Cc$ crystal structure of magnetite \cite{senn2012charge}. The two modes display a significant dependence on pump fluence. In particular, their energies soften dramatically towards the critical fluence ($F_C$) identified in previous studies \cite{de2013speed,randi2016phase} (Fig.~\ref{fig:Fig2}c). This behavior strikingly establishes their active involvement in the Verwey transition. In addition, the mode amplitudes first rise linearly until $\sim\,$1 mJ/cm$^2$ and then dramatically drop in the proximity of $F_C$ (Fig.~\ref{fig:Fig2}d), confirming the scenario of an impulsive Raman generation mechanism \cite{stevens2002coherent} that gets destabilized at higher fluences when the Verwey transition is approached due to transient lattice heating \cite{wall2012ultrafast,schaefer2014collective}. 

Finally, we establish whether the newly discovered collective excitations mutually interact. This is achieved by resolving the spectral region of the low-energy conductivity in which each coherent mode resonates. Figure~\ref{fig:Fig3}a shows the spectro-temporal evolution of the differential optical conductivity ($\Delta\sigma_1$) following photoexcitation. We observe the emergence of two broad features centered around 2 and 3 meV, which indicate how the lineshapes of the two modes in equilibrium are modified by the presence of the pump pulse. The shape and sign of the differential signal suggests that the main effect of this modulation is a broadening of the equilibrium Lorentzian lineshapes together with their amplitude change, though the exact form of the change is difficult to determine owing to the complexity of this material system. We then select a temporal trace at a representative THz photon energy (Fig.~\ref{fig:Fig3}b) and perform a Fourier transform analysis (Fig.~\ref{fig:Fig3}c). This yields the frequencies of the two collective modes expected from Fig.~\ref{fig:Fig2}a$-$c. Iterating this procedure at all photon THz energies (Fig.~\ref{fig:Fig3}d) reveals that the entire spectrum is modulated by both coherent electronic modes, demonstrating that these two modes are coupled to one another.

\begin{figure*}[t!]
\includegraphics[width=\textwidth]{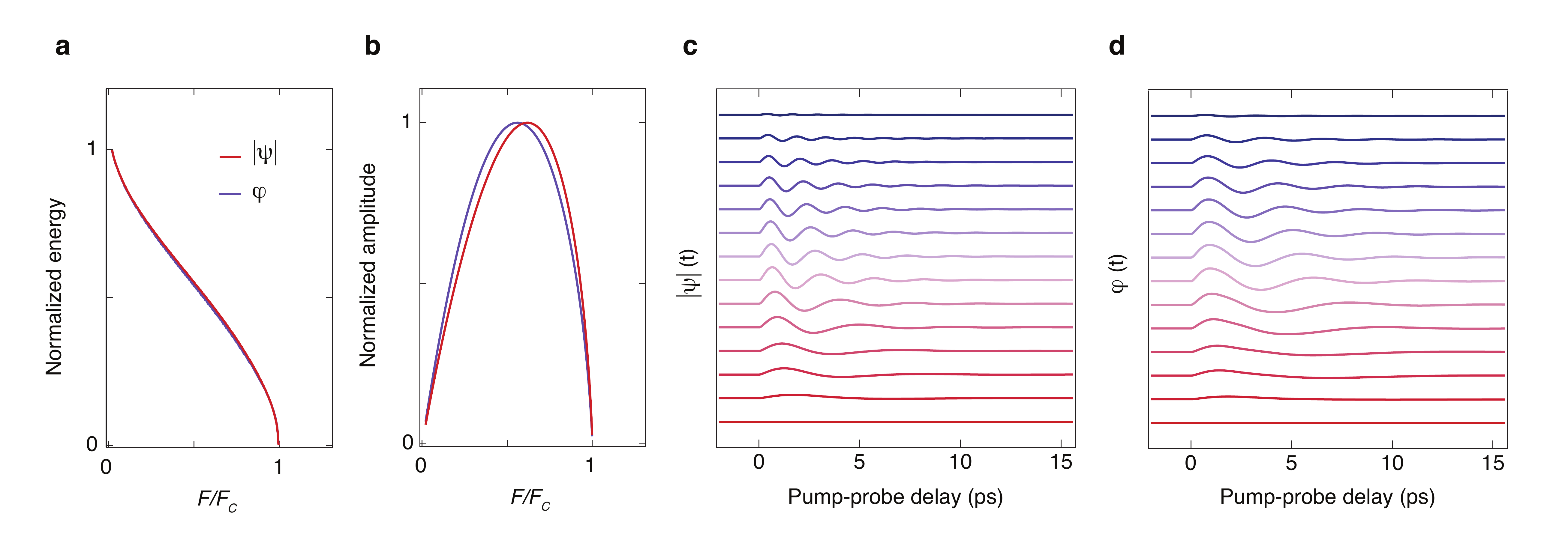}
\caption{\textbf{Time-dependent Ginzburg-Landau theory describing the dynamics of the collective modes.} \textbf{a},\textbf{b}, Dependence of the energies (\textbf{a}) and amplitudes (\textbf{b}) of both components of the order parameter ($|\psi|$ and $\varphi$) as a function of pump fluence. \textbf{c},\textbf{d}, Evolution of $|\psi|(t)$ (\textbf{c}) and $\varphi(t)$ (\textbf{d}) in the time domain over a range of fluences. The calculated behavior matches the qualitative trends observed in the experiments (see Fig.~\ref{fig:Fig2}b$-$d), namely the softening of the energies and the initial linear increase in the amplitudes followed by their destabilization towards the critical fluence $F_C$.}
\label{fig:Fig4}
\end{figure*}

To rationalize the behavior of these coherent collective modes after photoexcitation, we develop the simplest time-dependent Ginzburg-Landau (GL) model compatible with the symmetries of the system. In our calculations, we consider electronic collective mode fluctuations described by a complex order parameter $\psi = |\psi|e^{i\varphi}$. In the GL potential $F = F[\psi]$, we include a nonlinear term arising from electronic interactions, a linear coupling term between the real and imaginary parts of $\psi$ responsible for inversion symmetry breaking, and a pinning potential arising from impurity effects. The non-equilibrium action of the pump pulse on the mode fluctuations is introduced through a coupling to the intensity of the pump electric field. The full dynamics of the system are described by equations of motion that include phenomenological relaxation and inertial terms for both $|\psi|$ and $\varphi$ (see Methods). Despite its simplicity, our model captures the salient features of our experiment data. Specifically, the energies of both $|\psi|$ and $\varphi$ soften towards $F_C$ (Fig.~\ref{fig:Fig4}a) and the oscillation amplitudes of the modes rise linearly with increasing fluence before experiencing a dramatic quench in the proximity of $F_C$ (Fig.~\ref{fig:Fig4}b). Furthermore, there is a crossing of the amplitudes of the modes around 0.5$F_C$, which is also present in the experimental data (see Fig.~\ref{fig:Fig2}d). The resulting time dependences of $|\psi|$ and $\varphi$ are plotted in Figs.~\ref{fig:Fig4}c and~\ref{fig:Fig4}d, respectively, for several fluence values. While our model successfully reproduces the qualitative behavior of the experimentally observed dynamics, some quantitative mismatch is still present. In particular, deviations from the observed quasi-mean-field behavior of the mode energies in Fig.~\ref{fig:Fig2}c is due to the exact energy-fluence functional form used to describe the thermodynamic properties of the material.

Our results are rather surprising given the current understanding of magnetite's low-temperature phase. Since in this material the charge order is commensurate with the lattice and the Verwey transition is thought to be of an order-disorder type based on the observation of overdamped (i.e. diffusive) modes \cite{yamada1980neutron,bosak2014short,borroni2017coherent,borroni2018light}, the detection of underdamped (and therefore propagating) soft electronic modes is unexpected. This apparent contradiction can be reconciled by recalling that the trimeron order in magnetite leads to the development of a spontaneous ferroelectric polarization \cite{yamauchi2009ferroelectricity,senn2012charge,khomskii2014transition}. The ferroelectric instability is of the electronic (improper) type and involves charges that are weakly bound to the underlying lattice \cite{khomskii2014transition}. Low-energy modes can naturally emerge as collective fluctuations of charges within the trimeron network on top of the robust commensurate charge order. Though our observed modes seem reminiscent of the electronic component of amplitudons and phasons in conventional charge-density wave systems, such a simplified picture is not expected to capture the complexity of the trimeron order in magnetite. Here, we propose that these excitations can be described by oscillations of the trimeron network using a model of coherent polaron tunneling. In our calculations, we consider a quantum tunneling process involving self-trapped carriers on the central Fe$^{2+}$ sites of the trimerons and we compute the potential energy barrier for their coherent hopping to outer Fe$^{3+}$ sites using DFT \cite{yamauchi2009ferroelectricity}. By estimating the bonding-antibonding splitting arising from the superposition of the two states created by tunneling through the barrier, we find that its energy lies around 5 meV, i.e. within the monitored THz range (see Supplementary Note 1D). As shown schematically in Fig.~\ref{fig:Fig1}d, this coherent polaron oscillation corresponds to a sliding mode of the trimeron along its long axis. In analogy with nematicity, such motion is expected to create a smaller disturbance of the trimeron order compared to other alternatives such as the rotation of the $t_{2g}$ orbital participating in the trimeron (see Supplementary Note 1C for a DFT analysis of a scenario involving this type of orbiton). While at low temperature this energy splitting is robust and allows for coherent tunneling, as the temperature increases the polaron motion becomes overdamped, broadening the levels and reducing their relative splitting. As a result, the excitations soften with increasing temperature towards $T_V$ and are no longer supported in the absence of the trimeron order. We believe that only the development of advanced theoretical models will contribute to the identification of the real-space charge distribution characterizing these collective fluctuations. Indeed, their extremely low energy and small atomic displacements hinder their investigation with other steady-state \cite{huang2017jahn,elnaggar2018site} and time-resolved \cite{pontius2011time,de2013speed,pennacchio2018spatio} probes of charge/structural dynamics (see Supplementary Note 2). In contrast, our optical pump-THz probe approach enables the detection of these low-energy coherent modes with unprecedented sensitivity.

The current study highlights the strength of ultrafast THz probes in uncovering the soft character of electronic collective modes associated with an intricate order, in line with recent experiments on the Higgs and Leggett modes in superconductors \cite{matsunaga2014light,giorgianni2019leggett}. Beyond these results, we envision the use of strong THz fields \cite{kampfrath2013resonant} to resonantly drive the modes of the trimeron order in magnetite and similar charge-ordered compounds, enabling the coherent control of electronic ferroelectricity.

\section*{Methods}

\noindent\textbf{Single crystal growth and characterization.} A single crystal of synthetic magnetite oriented in the (111)-direction with a thickness of 0.5 mm was used in all experiments. The crystal was grown using the skull melting technique from 99.999\% purity Fe$_2$O$_3$. Afterwards, the crystal was annealed under a CO/CO$_2$ gas mixture to establish the appropriate iron-oxygen ratio. The quality and stoichiometry of the sample was characterized by measuring the AC magnetic susceptibility to determine the value of $T_V$. Supplementary Fig.~\ref{fig:FigS1} shows the real part of the AC magnetic susceptibility $\chi'$ as a function of temperature from 117 K to 126 K. The sudden decrease in $\chi'$ around 123 K indicates that $T_V \sim$ 123 K for this sample. This drop in the susceptibility results from microtwinning of the crystal when it undergoes its structural transition from cubic to monoclinic at $T_V$. In the low-temperature monoclinic phase, the ferroelastic domains constrain the motion of magnetic domain walls due to a higher cost in elastic energy. Consequently, the value of $\chi'$ should be lower in the monoclinic phase compared to the cubic phase \cite{balanda2005magnetic}. The temperature at which $\chi'$ exhibits this discontinuity therefore corresponds to $T_V$.\\ 

\noindent\textbf{Time-domain and ultrafast THz spectroscopy.} A Ti:Sapphire regenerative amplifier system with 100 fs pulses at a photon energy of 1.55 eV and a repetition rate of 5 kHz was used to generate THz pulses via optical rectification in a ZnTe crystal. The THz signal transmitted through the sample was detected by electro-optic sampling in a second ZnTe crystal with a 1.55 eV gate pulse. The frequency-dependent complex transmission coefficient was determined by comparing the measured THz electric field through the magnetite crystal to that through a reference aperture of the same size, and the complex optical parameters were then extracted numerically \cite{duvillaret1996reliable}.

For the ultrafast THz measurements, the output of the laser was split into a 1.55 eV pump beam and a THz probe beam, with the THz generation and detection scheme described above. The time delay between pump and probe and the time delay between the THz probe and the gate pulse could be varied independently. To measure the spectrally-integrated response, the THz time was fixed at the peak of the THz waveform and the pump-probe delay was scanned. The spectrally-resolved measurements were obtained by scanning both the THz time and the pump-probe delay time.\\

\noindent\textbf{Density functional theory calculations.} The crystal and electronic structure of the material was optimized using the projector augmented-wave method \cite{blochl1994projector} within the generalized gradient approximation \cite{perdew2008restoring} implemented in the VASP program \cite{kresse1996efficient}. The full relaxation of lattice parameters and atomic positions was performed in the crystallographic cell of the $Cc$ structure containing 224 atoms. The strong electron interactions in the Fe($3d$) states have been included within the local density approximation (LDA)+U method \cite{liechtenstein1995density} with the Coulomb interaction parameter $U=4.0$ eV and the Hund's exchange $J=0.8$ eV. The phonon dispersion curves were obtained using the direct method \cite{parlinski1997first} implemented in the Phonon software \cite{phonon2013}. The Hellmann-Feynman forces were calculated by displacing all non-equivalent 56 atoms from their equilibrium positions along the positive and negative $x$, $y$, and $z$ directions, and the force-constant matrix elements were obtained. The phonon dispersions along the high-symmetry directions in the first Brillouin zone were calculated by the diagonalization of the dynamical matrix. In the $Cc$ structure, there are 336 phonon modes at each wave vector. As discussed in the main text, Fig.~\ref{fig:FigEx1}a shows the calculated phonon dispersion curves in the low-energy range up to 20 meV. Fig.~\ref{fig:FigEx1}b displays the calculated partial phonon density of states projected on the Fe sites (black curve), which agrees well with experimental data taken from Ref. \cite{kolodziej2012nuclear} (red curve).\\

\noindent\textbf{Time-dependent Ginzburg-Landau calculations.} We constructed the simplest Ginzburg-Landau (GL) potential that captures phenomenologically the physics of the charge order in magnetite and is compatible with the symmetries of the system (see Supplementary Note 6). We defined the complex order parameter as $\psi = |\psi| e^{i \varphi}$, which is related to the real-space charge-density wave by $\delta \rho( \boldsymbol r) = {\rm Re} \{ \psi e^{i \boldsymbol q \cdot \boldsymbol r} \}$, where the wave vector could correspond to any linear combination of all the symmetry-allowed wave vectors. We modeled the transition as weakly first order, with a GL potential given by
\begin{equation}
F[|\psi|, \varphi] = \frac{a}{2}|\psi|^2 + \frac{b}{4} |\psi|^4 + \frac{d}{2} |\psi|^2 \sin \varphi  + \frac{g}{2} \cos \varphi + F_l,
\end{equation}
where $a(T) = -A(T_V-T)$ and $g(T) = -G(T_V-T)$ are functions of temperature and $b>0$ for stability. A non-zero amplitude-phase interaction term $d$ is allowed due to the lack of inversion symmetry in the low-temperature phase. The fourth term, proportional to the coefficient $g$, corresponds to a phenomenological ``restoring force'' that could arise from a pinning potential originating in short-range impurities \cite{fukuyama1978,lee1979,gruner1988,thomson2017}. This term could also emerge from a linear coupling with phonon modes belonging to the same irreducible representation as the charge modulation, where the proportionality constant $g$ would be a function of the electron-phonon coupling constant \cite{schafer2010,schaefer2014}. Finally, the coupling to the laser field was given by
$
 F_{l} = E(t)^2 (\eta_\psi |\psi|^2 + \eta_\varphi \varphi^2),
$
where $\eta_\psi$ and $\eta_\varphi$ are coupling constants. The pump electric field $E(t)$ was modeled as $E(t)^2 = 2 \mathcal F/(c \epsilon_0 T_p) \delta_{T_p}(t) T_p$, where $\mathcal F$ is the absorbed pump laser fluence, $c$ is the speed of light, $\epsilon_0$ is the permitivity of free space, $T_p \approx 0.1$ ps is the pump pulse duration, and $\delta_{T_p}(t)$ is a broadened delta function. The functional form used for the coupling to the laser field was chosen to mimic the force acting on the collective modes within the impulsive stimulated Raman scattering framework \cite{stevens2002coherent}.

Below $T_V$, $|\psi|^2$ acquires a finite expectation value denoted $\psi_{0}$ such that $
|\psi(t)|^2 =  1 + 2 \psi_0 |\delta \psi(t)| + O(\delta \psi(t) /\psi_0)^2.$ Due to the phase-amplitude mixing allowed by the lack of inversion symmetry, the laser also couples directly to the phase $\varphi$. In the dynamics, we included both relaxation and inertial terms for the amplitude and phase, and we allowed them to have different relaxation rates $\gamma_\psi$ and $\gamma_\varphi$. The effective temperature $T(t)$ is in general a function of time and laser fluence $\mathcal F$. However, due to the expected slow heat diffusion, as calculated in Supplementary Note 4, we assume that it remains at its initial effective value after photoexcitation during the whole measurement, $T(t)=T_f$ for $z<\lambda_p$. The differential equations governing the system were obtained by taking the variation of the GL potential with respect to the amplitude and phase around their equilibrium positions $\psi_0$ and $\varphi_0$. We obtained
\begin{align} 
\frac{\partial^2 |\psi|}{\partial t^2} & + \gamma_\psi \frac{\partial |\psi|}{\partial t} + a(T)|\psi|+ b|\psi|^3  + d |\psi| \varphi   = \eta_\psi E(t)^2 \psi_{0}, \label{eq:eq-motion_a} \\
 \frac{\partial^2 \varphi}{\partial t^2} & + \gamma_\varphi \frac{\partial \varphi}{\partial t} + d |\psi|^2 + g(T) \varphi = \eta_\varphi E(t)^2 \varphi_0 \label{eq:eq-motion_b}.  
\end{align}
We solved the coupled differential equations (\ref{eq:eq-motion_a}) and (\ref{eq:eq-motion_b}) numerically. The parameters used in Figs.~\ref{fig:Fig4}a--d are $A/b=0.13$ K$^{-1}$, $G/b=0.032$ K$^{-1}$, $\gamma_\psi/b=0.5$, $\gamma_\varphi/b=0.4$, $\eta_\varphi/\eta_\psi=0.505$, and $d/b=2.8$.

\section*{Data Availability} 
The data that support the findings of this study are available from the corresponding author upon reasonable request.
 
\section*{Acknowledgments}
Work at MIT was supported by the US Department of Energy, BES DMSE, Award number DE-FG02-08ER46521 and by the Gordon and Betty Moore Foundation's EPiQS Initiative grant GBMF4540. C.A.B. and E.B. acknowledge additional support from the National Science Foundation Graduate Research Fellowship under Grant No. 1122374 and the Swiss National Science Foundation under fellowships P2ELP2-172290 and P400P2-183842, respectively. M.R.-V. and G.A.F. were primarily supported under NSF MRSEC award DMR-1720595. G.A.F also acknowledges support from a Simons Fellowship. A.M.O. is grateful for the Alexander von Humboldt Foundation Fellowship (Humboldt-Forschungspreis). A.M.O. and P.P. acknowledge the support of Narodowe Centrum Nauki (NCN, National Science Centre, Poland),
Projects No. 2016/23/B/ST3/00839 and No. 2017/25/B/ST3/02586, respectively. D.L. acknowledges the project IT4Innovations National Supercomputing Center CZ.02.1.01/0.0/0.0/16$\_$013/0001791 and Grant No. 17-27790S of the Grant Agency of the Czech Republic. J.L. acknowledges financial support from Italian MAECI through the
collaborative project SUPERTOP-PGR04879, bilateral project AR17MO7,
Italian MIUR under the PRIN project Quantum2D, Grant No. 2017Z8TS5B,
and from Regione Lazio (L.R. 13/08) under project SIMAP.

\section*{Author contributions}
E.B. conceived the study. C.A.B., E.B., and I.O.O. performed the experiments. C.A.B. and E.B. analyzed the experimental data. A.K. grew the magnetite single crystals. P.P., D.L., K.P., and A.M.O. performed the density functional theory calculations. M.R.-V. and G.F. performed the time-dependent Ginzburg-Landau calculations. J.L. developed the model of coherent polaron tunneling with input from P.P. and contributed to the data interpretation. C.A.B., E.B., and N.G. wrote the manuscript with crucial input from all other authors. This project was supervised by N.G.

\section*{Competing interests}
The authors declare no competing interests.

\section*{Materials and correspondence}
Correspondence and requests for materials should be addressed to N.G.

\beginextdata
\begin{figure*}[t!]
\includegraphics[width=\columnwidth]{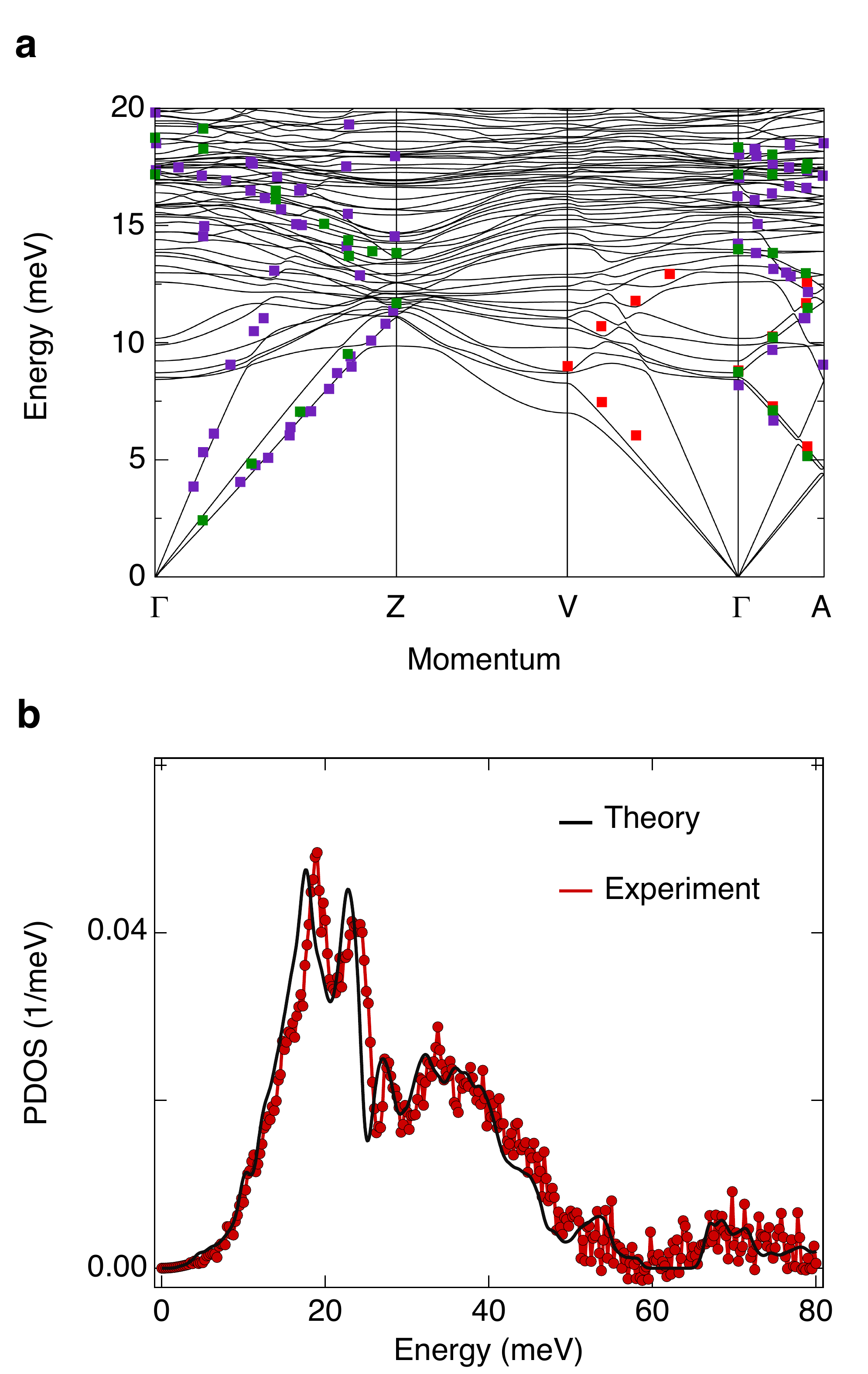}
\caption{\textbf{DFT calculation of the phonon dispersion in the $\boldsymbol{Cc}$ structure.} \textbf{a}, Low-energy phonon energy-momentum dispersion curves of magnetite calculated for the monoclinic $Cc$ symmetry. The symbols mark the energies of the phonon modes measured experimentally by inelastic neutron scattering (violet symbols from Ref. \cite{samuelsen1974low} and red symbols from Ref. \cite{borroni2017mapping}) and inelastic x-ray scattering (green symbols from Ref. \cite{hoesch2013anharmonicity}). There are no optical phonon branches in the energy range of the two newly-observed collective modes (1$-$4 meV). \textbf{b}, Partial phonon density of states projected on the Fe sites. The results of the DFT calculations are shown in black, while the experimental results at 50 K (taken from Ref. \cite{kolodziej2012nuclear}) are in red.}
\label{fig:FigEx1}
\end{figure*}

\clearpage
\beginsupplement
\onecolumngrid
\begin{center}
\textbf{\Large Supplementary Information for ``Discovery of the soft electronic modes of the trimeron order in magnetite''}
\end{center}\hfill\break
\twocolumngrid

\subsection{\normalsize Supplementary Note 1: Assignment of the collective modes}

In this section, we investigate the origin of the newly-discovered soft modes. We find that conventional collective modes such as phonons and magnons cannot account for our observations. We also consider theoretically a scenario in which orbital waves lead to a change in the trimeron direction at a given Fe$^{2+}$ site. Within DFT, we find that these orbital excitations lie at an energy much larger than the monitored terahertz range. Finally, we propose a theoretical model of coherent polaron tunneling of the low-temperature trimeron network that captures the energetics and behavior of our soft modes.

\begin{figure}[htb!]
\includegraphics[width=\columnwidth]{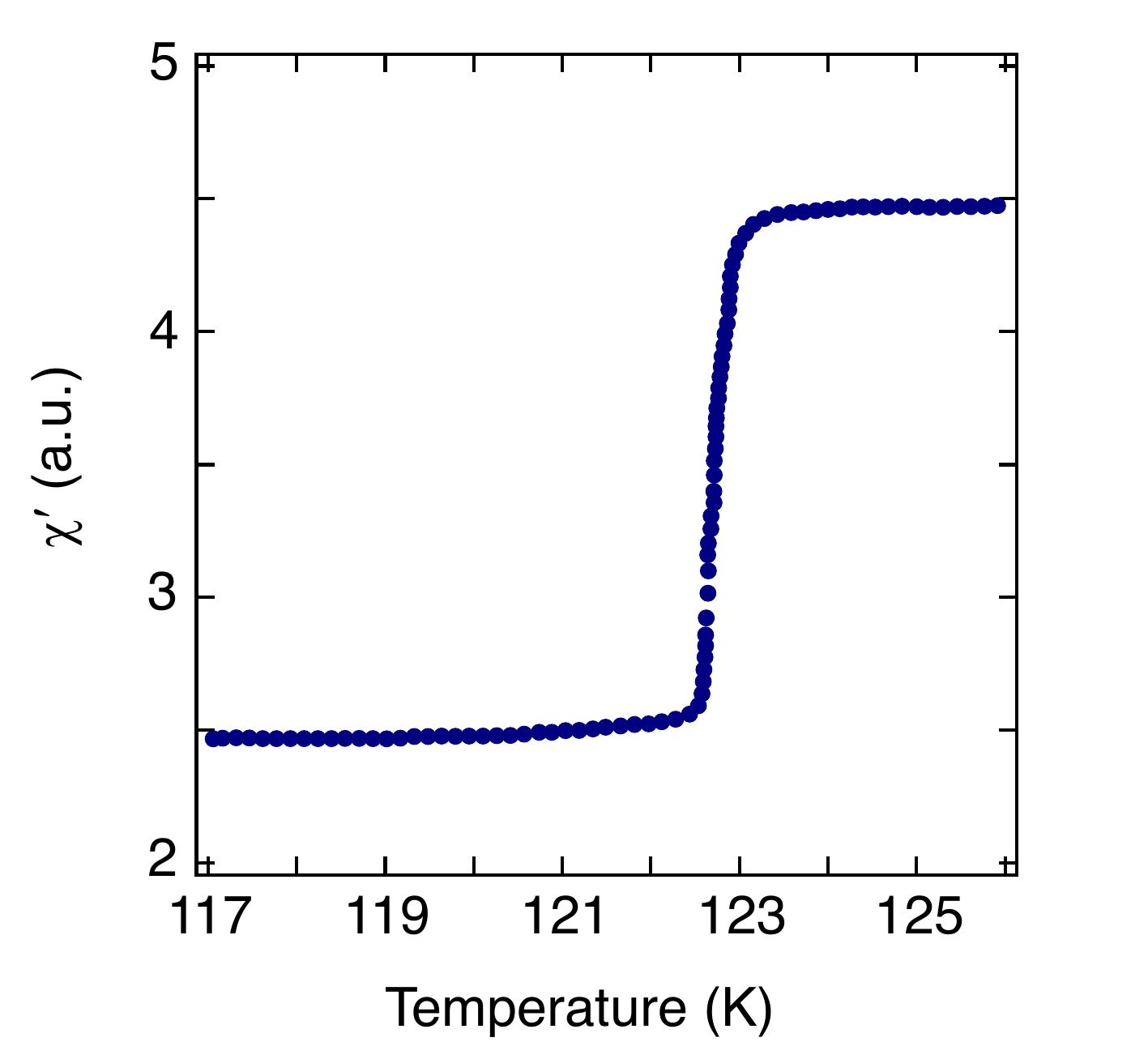}
\caption{\textbf{AC magnetic susceptibility of our magnetite crystal.} Real part of the AC magnetic susceptibility $\chi'$ as a function of temperature. From the temperature at which $\chi'$ undergoes a sudden drop, we obtain $T_V \sim$ 123 K for this crystal.}
\label{fig:FigS1}
\end{figure}

\subsubsection{\normalsize A. Phonons}

First, we rule out the possibility that the collective modes observed in equilibrium (Fig.~\ref{fig:Fig2}a in the main text) are due to optical phonons or folded acoustic phonons by performing \textit{ab initio} calculations of the phonon dispersion in the $Cc$ structure of magnetite (Fig.~\ref{fig:FigEx1}a). The calculated dispersion reveals that there are no optical phonon branches at the $\Gamma$ point in the energy range corresponding to the two observed modes. Folded acoustic phonons (obtained by a folding of the Brillouin zone) likewise have much higher energies than our two modes. 

For the oscillations in the time domain, their energies are very close to those of the two collective modes in equilibrium at 7 K, indicating that they are indeed the same modes. In this case, we also rule out any alternative explanation associated with phonons. It is important to consider the possibility that the observed oscillations in the time domain are coherent acoustic phonons (CAPs) generated by the pump pulse through the deformation potential, the thermoelastic coupling, and the inverse piezoelectric effect \cite{ruello2015physical}. The absence of any dispersion in the spectrally-resolved data excludes this scenario. Indeed, when the probe photon energy is tuned in a spectral range where the material is highly transparent (i.e. typically below the fundamental optical gap), the Brillouin scattering condition $\lambda_{\text{phonon}}=\frac{\lambda_{\text{probe}}}{2n}$ holds \cite{brillouin1922diffusion}. If the refractive index $n$ is roughly constant across the probed range (as is the case here -- see Supplementary Note 3A), then the wavelength of the coherent strain should depend on the probe photon energy. Since we do not observe any variation in the frequency of the oscillations at different probe photon energies (see Fig.~\ref{fig:Fig3} in the main text), the oscillations cannot be CAPs. Furthermore, CAPs typically exhibit a cosine behavior as a function of time (due to the nature of the generation process \cite{ruello2015physical}), whereas the oscillations in our experiment are closer to sine functions. The latter is indicative of a scenario in which the force acting on the collective modes is more impulsive than displacive in nature \cite{stevens2002coherent}.

\subsubsection{\normalsize B. Magnons}

In addition to the discussion of magnons in the main text, we note that folded acoustic magnetic modes can be ruled out as well due to their energies not falling in our energy range \cite{mcqueeney2005influence}. Also, the frequency of the spontaneous ferromagnetic resonance in magnetite is $\sim\,$16 GHz \cite{bickford1950ferromagnetic}, which is extremely low. Recent pump-probe magneto-optical measurements under our photoexcitation conditions revealed the slow coherent oscillations of this ferromagnetic resonance mode \cite{panigrahi2019spins}. Despite a high time resolution of 50 fs, no magnetic modes with energies close to our observed soft excitations were detected. This further confirms that our newly-discovered soft modes are non-magnetic in origin.

\begin{figure}[htb!]
\includegraphics[width=\columnwidth]{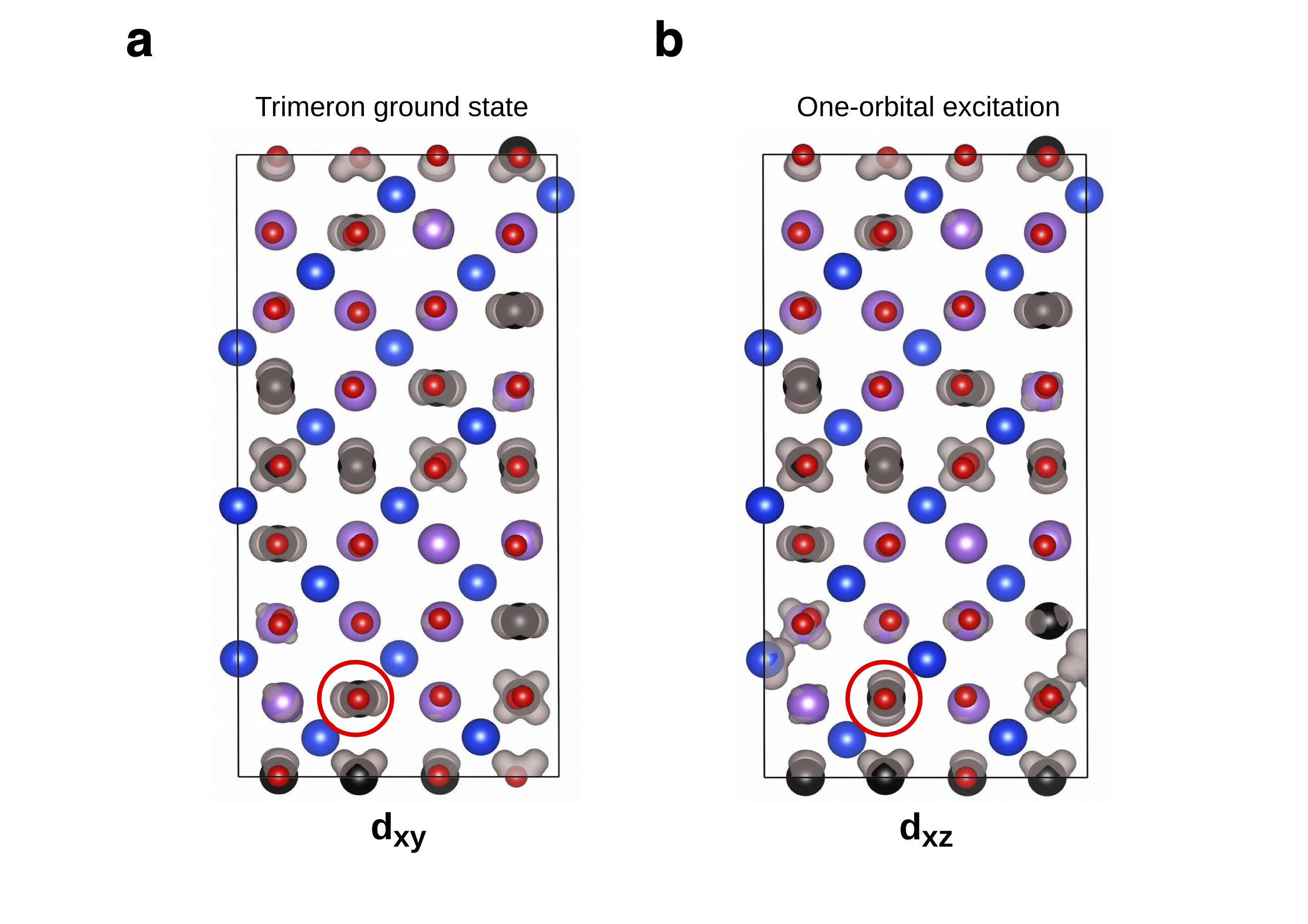}
\caption{\textbf{DFT results on the one-orbital excitation.} Maps of the minority magnetization highlighting the orbital occupation of $t_{2g}$ states (displayed as gray shapes). The black, purple, and red spheres represent Fe$^{2+}$-B sites, Fe$^{3+}$-B sites, and O sites, respectively. Blue spheres represent Fe$^{3+}$-A sites. Panel \textbf{a} presents the trimeron ground state, whereas panel \textbf{b} shows the one-orbital excitation stabilized in our calculations. We observe that the orientation of the orbital indicated by the red circle is rotated and the charge distribution around it is modified.}
\label{fig:FigS2}
\end{figure}

\subsubsection{\normalsize C. Orbitons}

As shown in Fig.~\ref{fig:Fig1}b in the main text, a trimeron consists of a central Fe$^{2+}$-B site with one $t_{2g}$ orbital occupied. Two lobes of the $t_{2g}$ orbital point towards nearest neighbor Fe$^{3+}$-B sites, which are the outer sites of the trimeron. One elementary excitation that can emerge involves a $d$-$d$ transition within the central B site. As a result of the orbital change, the orientation of the trimeron is rotated. In our study, we looked for such metastable configurations with one rotated trimeron in the unit cell using DFT calculations. We found that it is indeed possible to stabilize this metastable state if a proper initial state is chosen in the DFT minimization. To create the initial state, we temporarily reduced the Hubbard $U$ in an Fe$^{2+}$ site until in-gap states were formed. Then, we inverted the population of the states closer to the Fermi level and relaxed the charge and the lattice. Finally, we restored the physical $U$ and the filling of the states according to the Aufbau rule, relaxing again the charge and lattice degrees of freedom. We found that a metastable state with a rotated orientation can form. Figure ~\ref{fig:FigS2} shows the calculated minority magnetization, highlighting the orbital occupation of the $t_{2g}$ states. Figure~\ref{fig:FigS2}a presents the trimeron ground state, whereas Fig.~\ref{fig:FigS2}b shows the metastable (one-orbital excitation) state. We see that the orientation of the orbital indicated by the red circle is rotated. Due to the low symmetry of the problem and the large number of atoms in the $Cc$ unit cell (224 atoms), such computations are very expensive and it is not feasible to exhaust all possible trimeron sites and orientations. For the studied cases, we find that the theoretical energy for these excitations lies on a scale of $\sim\,$1 eV, which is much larger than the energy of the experimentally-observed excitations (a few meV). The large energy cost involved in the one-orbital excitation is due to the disruption of the trimeron network. All the trimerons are linked by their outer sites (see Fig.~\ref{fig:Fig1}a in the main text) and the modification of one trimeron orientation significantly influences the charge-orbital occupations of the neighboring atoms, as indicated in Fig.~\ref{fig:FigS2}b.  While there may be less disruptive excitations with respect to those studied, we think it is unlikely due to the large energy cost found in the investigated configurations.

\subsubsection{\normalsize D. Polarons}

\begin{figure}[htb!]
\includegraphics[width=0.8\columnwidth]{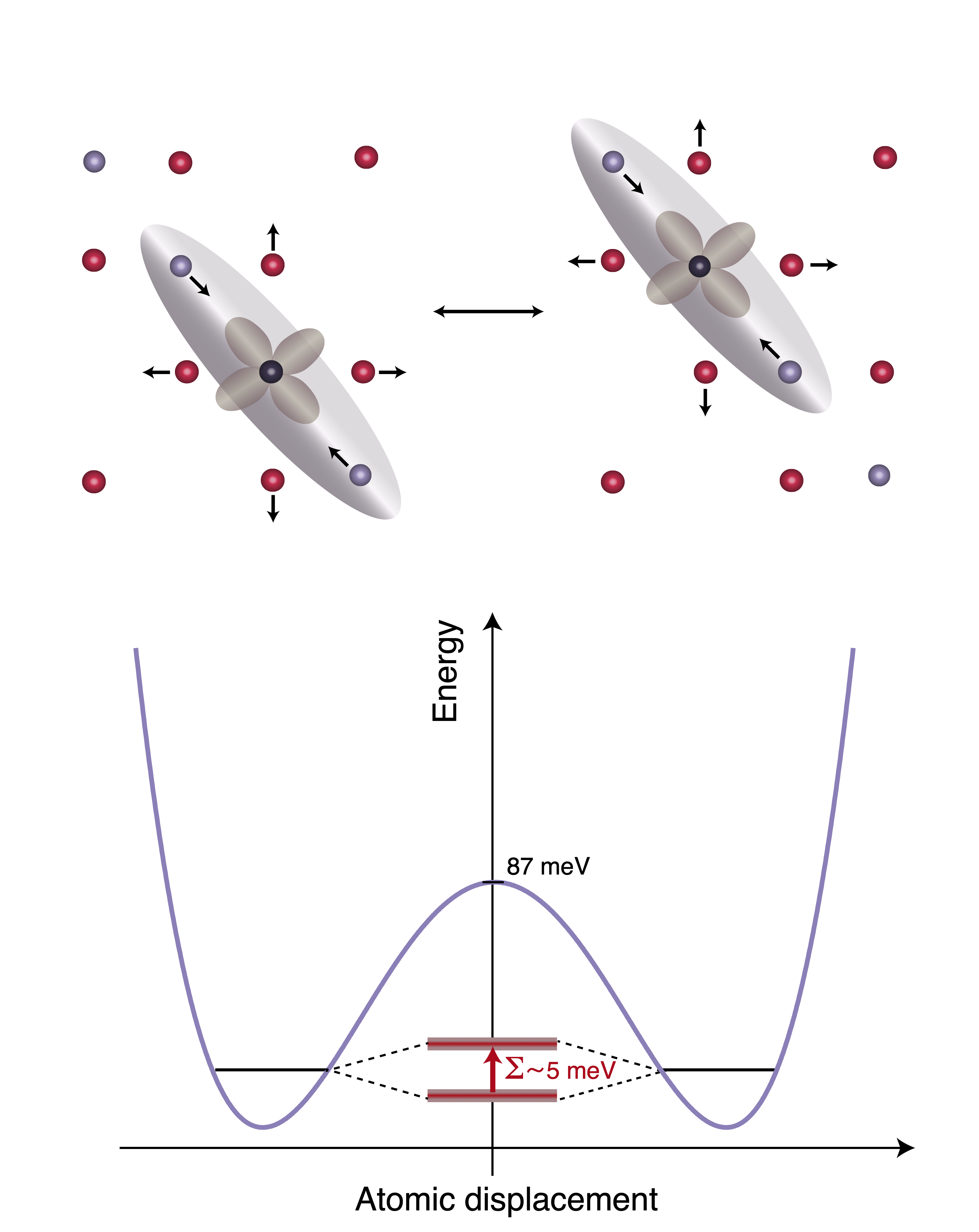}
\caption{\textbf{Schematic representation of the coherent polaron tunneling.} The cartoon shows the process of a polaron tunneling from an Fe$^{2+}$ site to a neighboring Fe$^{3+}$ site (for simplicity, we display only a single trimeron). The purple, black, and red spheres represent Fe$^{3+}$, Fe$^{2+}$, and O sites, respectively. The atomic displacements involved in the polaron formation have been adapted from Ref. \cite{yamauchi2009ferroelectricity}. The single polaron barrier (violet curve) is computed with DFT and all polarons are assumed to move coherently to the nearest neighbor site along the crystallographic $a$-axis. The height of the barrier is determined by symmetry from the energy required to displace the ions from a nonpolar configuration to the ferroelectric \textit{Cc} structure. In this model, the excitations observed experimentally (represented as a red arrow) stem from the bonding-antibonding splitting $\Sigma$, whose energy scale of $\sim\,$5 meV is exaggerated in the drawing compared to the potential barrier for visualization purposes.}
\label{fig:FigS3}
\end{figure}

Finally, we argue that a plausible explanation for the existence of the newly-discovered excitations involves tunneling of polarons that give rise to the electrical polarization. Specifically, we assume that a polaron can tunnel from one Fe site to the neighboring one with a matrix element $t_\text{eff}$. In this model, the excitations observed experimentally stem from the bonding-antibonding splitting $\Sigma$ = 2$t_\text{eff}$. From Fig. 3 of Ref. \cite{yamauchi2009ferroelectricity} we observe that in the primitive cell there are 16 sites of the Fe$^{2+}$ type. The number of Fe$^{2+}$ atoms in the different layers of the primitive cell (from layer 0/8 to 7/8) is 4, 2, 1, 1, 4, 2, 1, 1. It is natural to assume that the lowest-energy excitation comes from the layers with one Fe$^{2+}$ being the core of a trimeron which can tunnel to a neighboring Fe$^{3+}$ without being hindered by other trimerons. Such a process can arise along the $a$ direction (layers 2/8 and 6/8) or along the $b$ direction (layers 3/8 and 7/8). As a proxy to the single polaron barrier, we compute within DFT the barrier in which all four polarons move coherently to the next nearest neighbor site along the shortest Fe-Fe distance. This effectively produces a $C_2$-rotated image of the structure with the height of the barrier determined by symmetry from the energy required to displace the ions from a nonpolar configuration to the ferroelectric \textit{Cc} structure. A schematic representation of the coherent polaron tunneling mechanism is shown in Fig.~\ref{fig:FigS3}. We obtain that the barrier height is 87 meV per polaron, which agrees with the value that can be deduced from previous studies \cite{yamauchi2009ferroelectricity}. Next, we interpret this as the binding energy of a polaron ($E_p$) in a Holstein-like model. Using a Lang-Firsov transformation \cite{lang1963kinetic, ranninger1992two}, we can estimate the matrix element for coherent polaron hopping as $t_\text{eff}$ = $t$ $e^{-E_p/\hbar\omega_0}$, where $t$ is the bare electron hopping integral and $\hbar\omega_0$ is the effective phonon energy.  We evaluate the energy scale associated with this process by substituting the relevant parameters for magnetite. The main displacements to stabilize the polaron are due to the oxygen atoms surrounding the polaron sites. Projecting the phonon density of states on these polaron Fe sites, we find a low-energy peak around 20 meV (Fig.~\ref{fig:FigEx1}b). We interpret this as the  characteristic phonon energy $\hbar\omega_0$. We extract the value of the direct Fe $d$-$d$ hopping from our DFT calculations, which yield a value of $t$ $\sim$ 200 meV. The latter is in agreement with a simple estimate that can be given by the Harrison method \cite{harrison2012electronic}, considering an Fe-Fe distance of 2.97 \AA. As a result, we obtain $\Sigma$ $\sim$ 5.2 meV, a value that lies on the same energy scale of the modes observed experimentally. One possible explanation for the observation of two low-energy modes in our experiments could be polaron tunneling in the two non-equivalent directions, $a$ and $b$. 
At low temperatures the trimerons form a crystal and they cannot diffuse, i.e. they are confined in the insulating state. The identified jumps are likely the lowest energy allowed in the trimeron lattice. As the temperature is raised, the confinement potential weakens and the polaron jumps become the precursor of the Drude peak \cite{lorenzana2001instability}. Thus, it is natural that the excitations soften and become overdamped as observed experimentally.

\subsection{\normalsize Supplementary Note 2: Detectability of the soft modes through other experimental probes}

In this section, we consider other possible experimental techniques, in particular structural probes, that might be able to observe these newly-detected electronic collective modes in magnetite. We find that the current structural probes either have intrinsic limitations in detecting these modes (in terms of sensitivity or monitored spectral range) or cannot provide additional information about their nature.

In equilibrium, high-resolution spontaneous Raman scattering \cite{takesada2006perfect} would be able to observe these low-energy modes (2 and 4 meV) as they are also Raman active. In this technique, the lower bound of the energy for the detection of collective modes is $\sim\,$0.12 meV. However, a Raman study would not provide any additional information about the nature of our modes beyond what we observe with our impulsive Raman-type  measurement. One can also perform time-resolved Raman scattering \cite{versteeg2018tunable,kukura2007femtosecond}, but this technique poses severe constraints on the detection of our low-energy collective modes, namely: (i) the energy-time resolution in the spontaneous version of this method due to its Fourier transform-limited nature; (ii) the presence of the elastic (Rayleigh) peak in the spontaneous version; (iii) the scattering from the fundamental laser beam into the spectrometer in the stimulated version of this technique. Similar to the equilibrium case, even if all of these obstacles were overcome, time-resolved Raman scattering would not provide significant information beyond what can be extracted from our experiment. To shed light on the properties of these electronic modes, one should determine their energy-momentum dispersion relation. While in principle this is possible through resonant inelastic x-ray scattering (RIXS) or electron energy loss spectroscopy (EELS), in practice these methods lack sufficient energy resolution. Low-energy modes around 2--4 meV would be obscured by the broadening (at least 10 meV) of the elastic peak at zero energy \cite{kogar2017signatures}, hindering the extraction of any meaningful information. Similarly, the time-resolved versions of these techniques \cite{cao2019ultrafast,pomarico2018ultrafast} are limited by the achievable energy resolution in the presence of ultrashort laser pulses.

Since our modes contain both an electronic and a structural component, they should in principle be observable in time-resolved structural experiments. These probes can circumvent the energy resolution constraints of most steady-state techniques by revealing the modes as coherent oscillations in the time domain (similar to Fig.~\ref{fig:Fig2}b in the main text) in response to photoexcitation. However, time-resolved structural measurements using our same experimental conditions (temperature and fluence range, pump photon energy, and time resolution) have already been reported \cite{pontius2011time,de2013speed,pennacchio2018spatio} and no coherent oscillations were observed. Indeed, the sensitivity currently offered by these methods is orders of magnitude below what is needed to detect the coherent oscillations of our modes (the change in the terahertz electric field due to our coherent modes is on the order of 10$^{-5}$). Therefore, our terahertz detection setup provides a tool to examine low-energy coherent modes with unprecedented sensitivity.

\subsection{\normalsize Supplementary Note 3: Additional equilibrium data}

\subsubsection{\normalsize A. Refractive index at 7 K}

\begin{figure}[htb!]
\includegraphics[width=0.8\columnwidth]{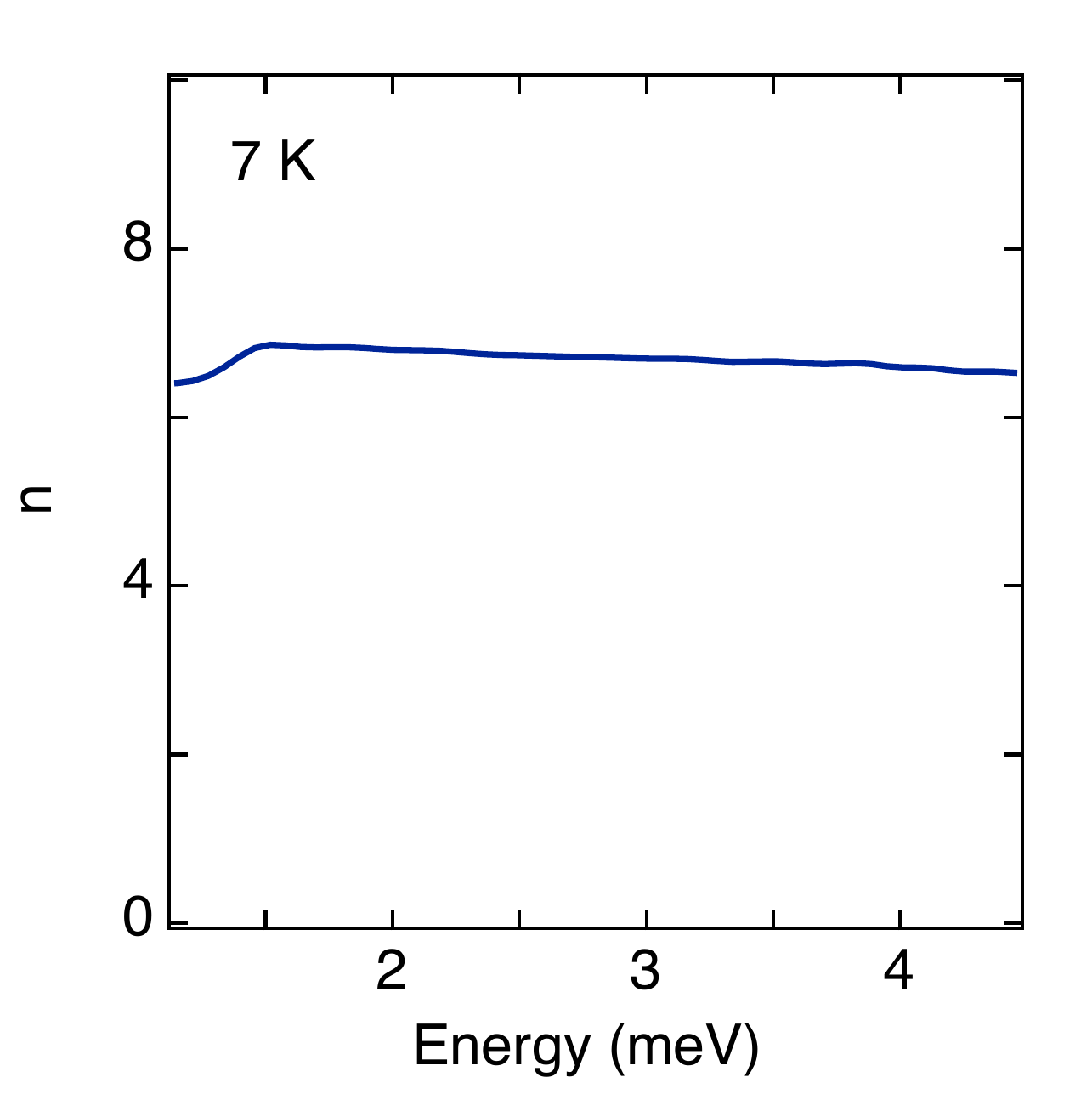}
\caption{\textbf{Refractive index of magnetite at low temperature.} Real part of the refractive index ($n$) measured by time-domain terahertz spectroscopy at 7 K. There is almost no variation in the index over this energy range.}
\label{fig:FigS4}
\end{figure}

The refractive index was extracted from the equilibrium terahertz data as described in the Methods section. Figure~\ref{fig:FigS4} displays the real part of the refractive index ($n$) in the terahertz region shown in Fig.~\ref{fig:Fig3} in the main text. The index exhibits very little variation across this energy range and can therefore be approximated as constant.

\subsubsection{\normalsize B. Equilibrium optical conductivity at 100 K}

\begin{figure}[htb!]
\includegraphics[width=0.8\columnwidth]{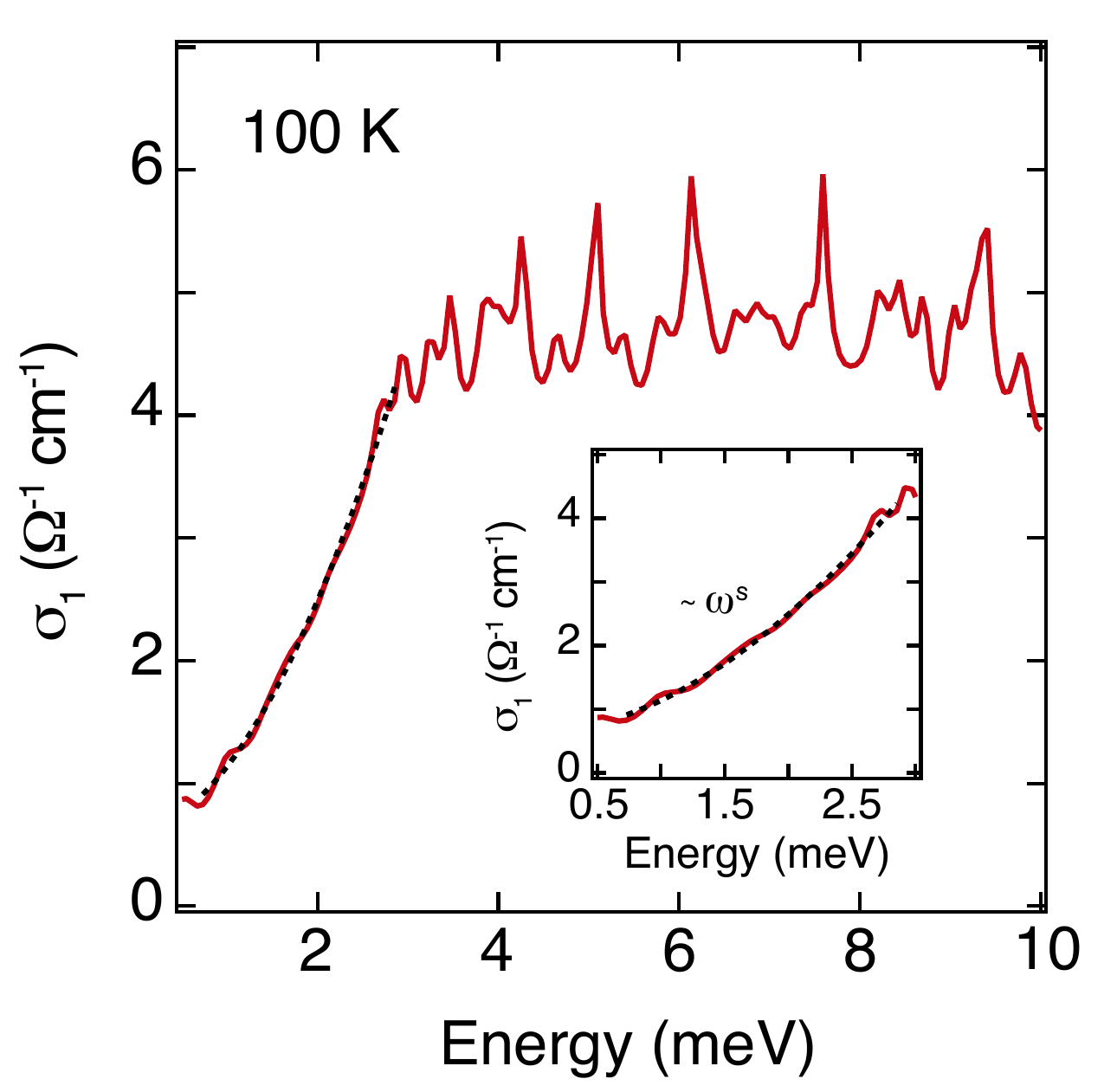}
\caption{\textbf{Power-law dependence of the low-energy optical conductivity just below $\boldsymbol{T_V}$.} Real part of the optical conductivity ($\sigma_1$) in the terahertz range at 100 K from Fig.~\ref{fig:Fig2}a in the main text. At low energies (inset), the curve is fit to a power law with exponent $s$ = 1.8, in close agreement with previously reported terahertz measurements \cite{pimenov2005terahertz}.}
\label{fig:FigS5}
\end{figure}

As discussed in the main text, our equilibrium optical conductivity data at 100 K agrees very well with previously reported measurements of the terahertz conductivity of magnetite in this temperature regime, in which a power-law dependence is observed and attributed to hopping between localized states \cite{pimenov2005terahertz}. Figure~\ref{fig:FigS5} shows the real part of the optical conductivity ($\sigma_1$) at 100 K and its fit to a power-law function of the form $\sigma_1 \sim \omega^s$ (for simplicity, we neglect the additional factor accounting for the saturation at higher frequencies used in Ref. \cite{pimenov2005terahertz} and instead focus only on the low-frequency regime). The power-law fit (black dashed line) yields $s=1.8$, which is similar to the value of $s=1.5$ reported in Ref. \cite{pimenov2005terahertz} for 100 K. This provides a quantitative measure of the close agreement between the two data sets.

\subsection{\normalsize Supplementary Note 4: Transient effective lattice temperature and heat diffusion}

\begin{figure*}[htb!]
\includegraphics[width=0.8\textwidth]{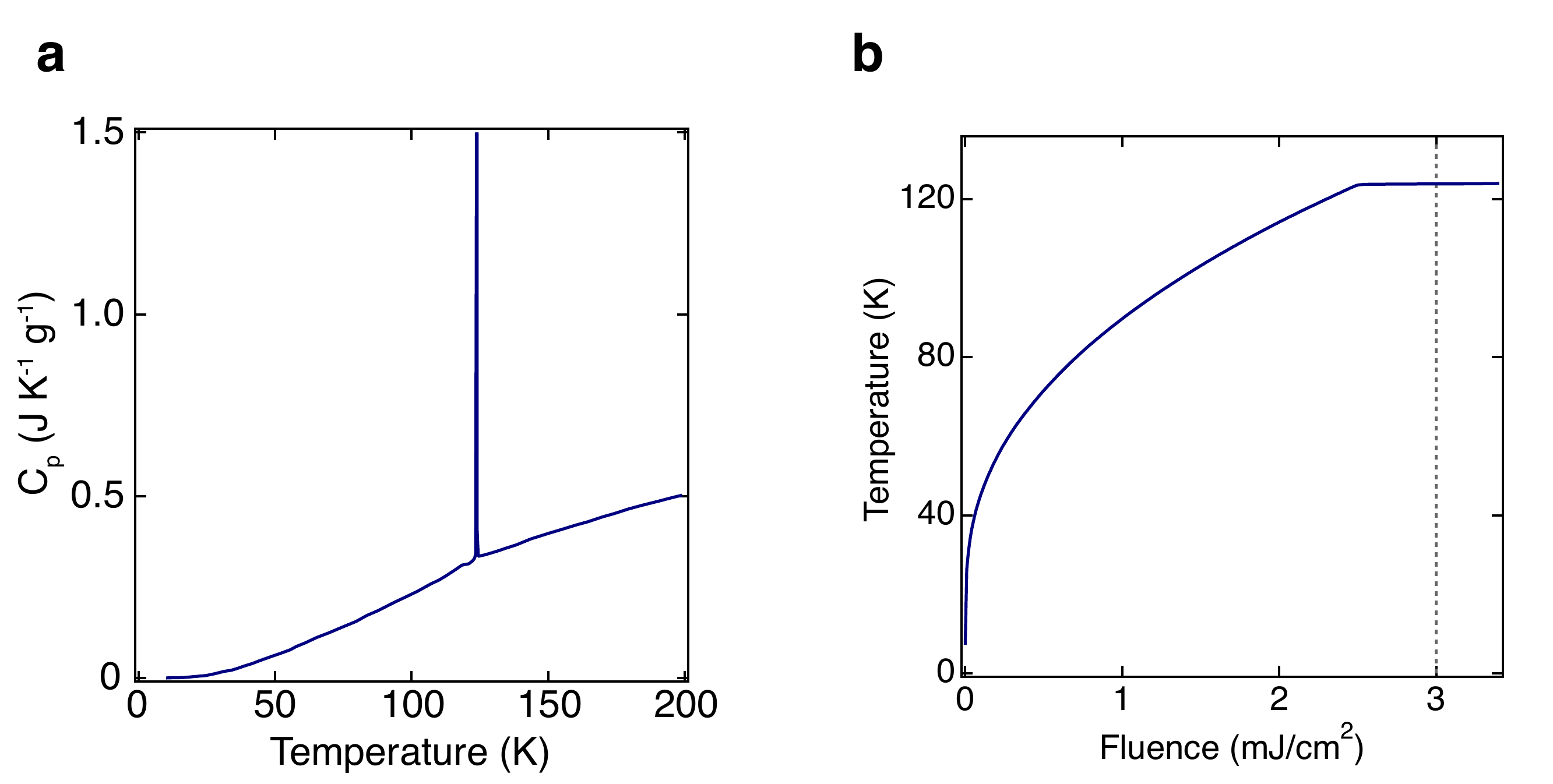}
\caption{\textbf{Relationship between pump laser fluence and effective sample temperature.} \textbf{a}, Heat capacity data used in the calculation of the effective temperature taken from Ref. \cite{takai1994}. \textbf{b}, Effective temperature as a function of laser fluence obtained by solving Eq.~(\ref{eq:fluence_v_T}). The dashed line indicates the maximum fluence used in the experiments.}
\label{fig:FigS6}
\end{figure*}

In this section, we connect the pump laser fluence ($\mathcal{F}_0$) with a rise in the effective lattice temperature ($T$) after photoexcitation. The fluence effectively deposited into the sample is
\begin{equation}
\mathcal{F} = \mathcal{F}_0 (1-R) (1-e^{-z/\lambda_p}),
\end{equation}
where $R$ is the sample reflectivity, $\lambda_p = 330$ nm is the penetration depth for $1.55$ eV photons, and $z$ labels the direction of light propagation. Energy conservation imposes a relation between fluence and temperature,
\begin{equation}
\mathcal{F} = \frac{m}{A} \int_{T_i}^{T_f} C_p(T) dT,
\label{eq:fluence_v_T}
\end{equation}
where $A$ and $m$ are the sample area and mass (determined by the penetration depth $\lambda_p$) illuminated by the pump laser, respectively, $T_i$ = 7 K is the temperature of the sample prior to the pump pulse arrival, $T_f$ is the effective temperature reached after photoexcitation, and $C_p(T)$ is the temperature-dependent heat capacity at constant pressure. In a solid, due to the negligible change in volume during the heating process, the heat capacity at constant pressure and at constant volume are expected to be very similar. 

Equation~(\ref{eq:fluence_v_T}) can be solved numerically for $T_f$. Using data of the heat capacity of a synthetic magnetite crystal taken from Ref. \cite{takai1994} (Fig.~\ref{fig:FigS6}a), we obtain $T_f$ as a function of fluence (Fig.~\ref{fig:FigS6}b). According to this estimate, the effective temperature after photoexcitation rises to 124 K for the maximum fluence used in the experiment ($\mathcal{F}_{max}$ = 3.0 mJ/cm$^2$), a temperature that is approximately equal to $T_V$. 

\begin{figure}[htb!]
\includegraphics[width=0.8\columnwidth]{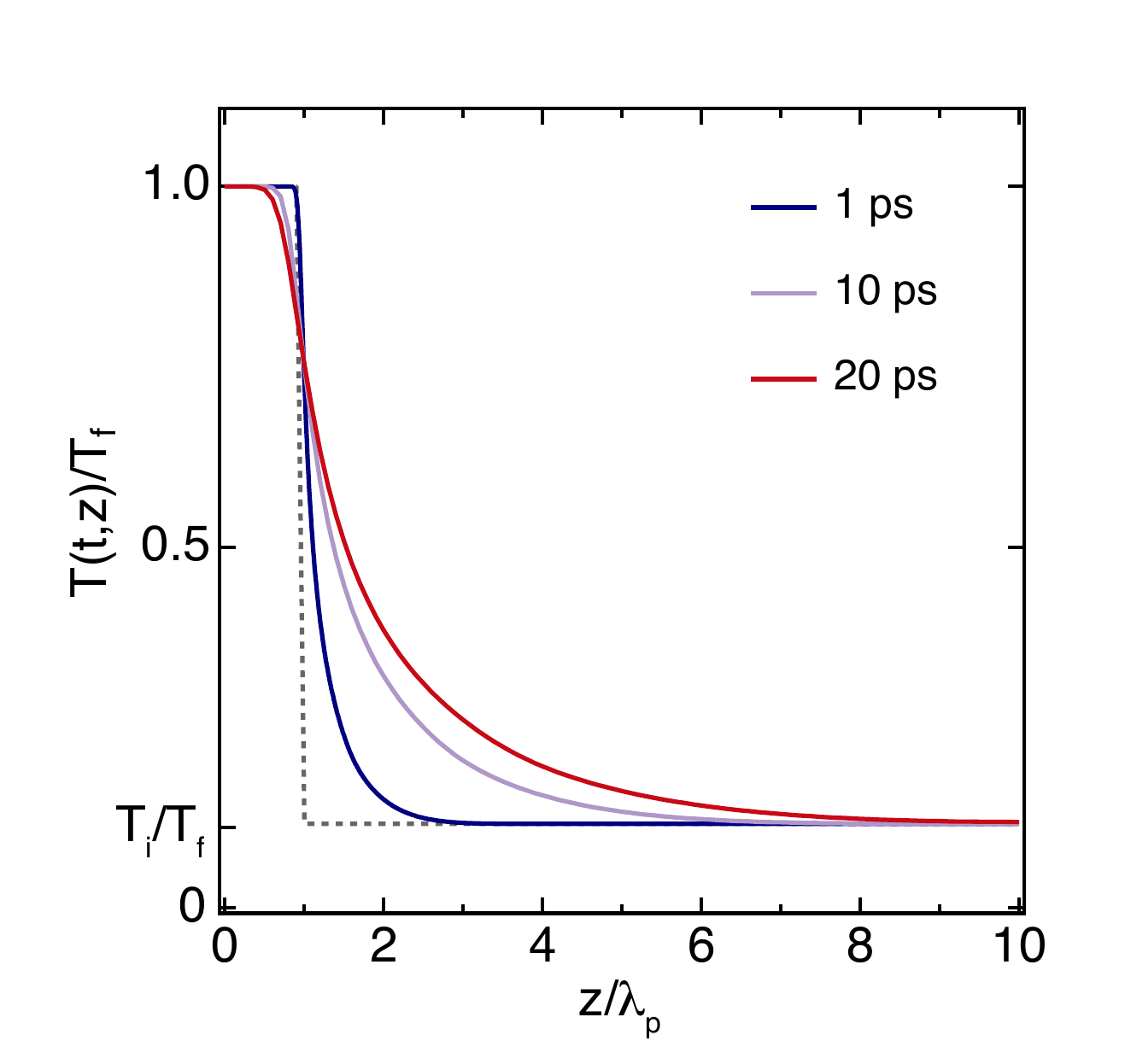}
\caption{\textbf{Heat diffusion through the sample.} Effective temperature as a function of position $z$ in the sample at three different times. The gray dashed line corresponds to the temperature profile immediately after photoexcitation. In these calculations the full diffusion equation is solved, taking into account the temperature dependence of the heat capacity at constant pressure $C_p$ and the thermal conductivity $\kappa$. The penetration depth is $\lambda_p = 330$~nm, while the sample depth is $L$ = 0.06 cm.}
\label{fig:FigS7}
\end{figure}

The time evolution of the transient lattice temperature is given by the heat diffusion equation, 
\begin{equation}
 \partial_t \left(\rho  \int^{T(t,z)}_{T_i} C_p(\mathcal T) d\mathcal{ T} \right) = \nabla \cdot \left( \kappa(T) \nabla T(t,z) \right),
 \label{eq:heat_eqn}
\end{equation}
subject to the initial condition $T(0,z)=T_{f}$ if $z < \lambda_p$ and $T(0,z)=T_i$ = 7 K otherwise, and insulating boundary conditions $\partial_z T(t,z)|_{z=0,L} =0$. In Eq.~(\ref{eq:heat_eqn}), $\rho$ = 5.175 g/cm$^3$ is the sample density and $\kappa$ is the temperature-dependent thermal conductivity. Given that the sample is nearly uniformly illuminated along the plane, we consider only the spatial dimension along the depth of the sample. We solve Eq.~(\ref{eq:heat_eqn}) for $T(z,t)$. The results, shown in Fig.~\ref{fig:FigS7}, demonstrate that the effective temperature changes only slightly in the vicinity of the photoexcited region in the time scale relevant for the experiments. 

\begin{figure*}[htb!]
\includegraphics[width=0.8\textwidth]{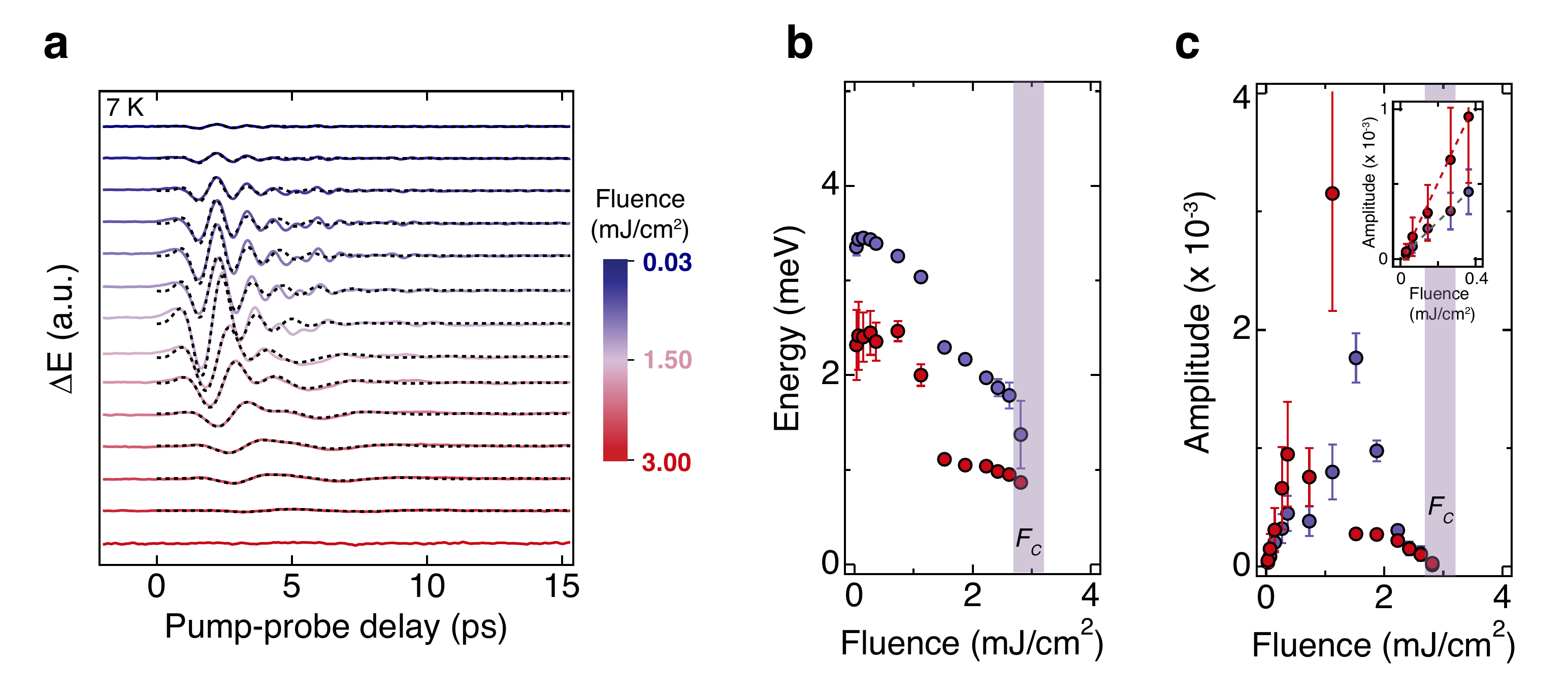}
\caption{\textbf{Second data set of the pump fluence dependence.} \textbf{a}, Second data set of the pump-induced change in the terahertz electric field ($\Delta E$) transmitted through the sample at 7 K following photoexcitation at various absorbed fluences. The traces are offset vertically for clarity. Each curve was fit to two damped sine waves and the fits in the time domain are displayed as dashed black lines. \textbf{b}, Energy of each oscillation extracted from the fits as a function of fluence. \textbf{c}, Amplitude of each mode versus fluence. The inset shows the linear rise of the amplitudes in the low fluence regime, which is compatible with an impulsive Raman excitation process. The shaded bars in \textbf{b} and \textbf{c} indicate the critical fluence for melting the trimeron order as reported in Refs. \cite{de2013speed,randi2016phase}. The error bars in \textbf{b} and \textbf{c} represent the 95\% confidence interval for the corresponding fit parameters.}
\label{fig:FigS8}
\end{figure*}

\begin{figure*}[htb!]
\includegraphics[width=0.8\textwidth]{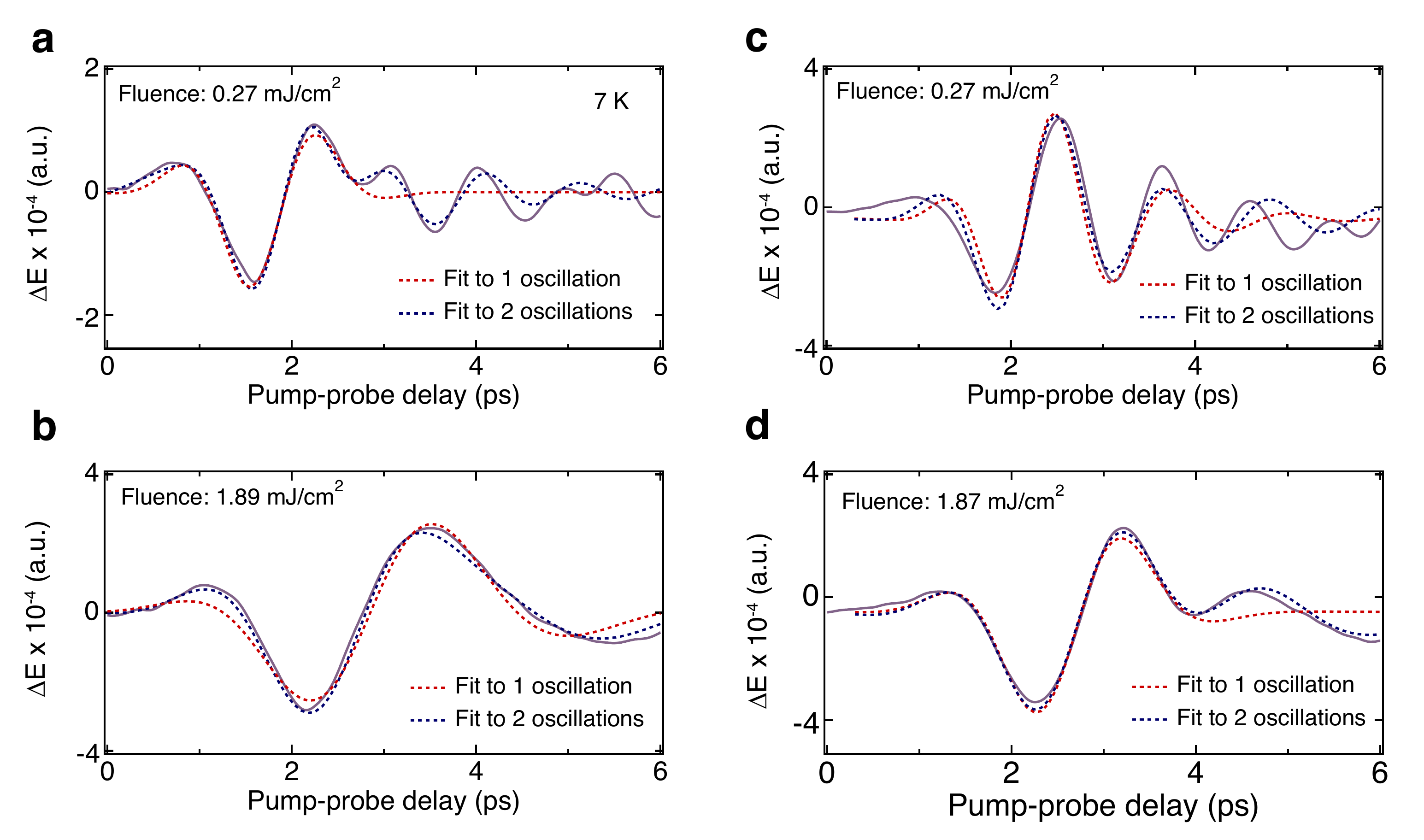}
\caption{\textbf{Fits of the pump-probe response in the time domain.} \textbf{a},\textbf{b}, Pump-probe response at low fluence (\textbf{a}) and high fluence (\textbf{b}) for the data set in Fig.~\ref{fig:Fig2}b in the main text. \textbf{c},\textbf{d}, Pump-probe response at low fluence (\textbf{c}) and high fluence (\textbf{d}) for the data set in Fig.~\ref{fig:FigS8}a. The fits to one (red) and two (blue) damped sine waves are shown for each curve. A single oscillation is unable to capture all the features of the data, demonstrating that two frequencies, and therefore two coherent modes, are present at all fluences.}
\label{fig:FigS9}
\end{figure*}

\subsection{\normalsize Supplementary Note 5: Additional pump-probe data and analysis}

\subsubsection{\normalsize A. Second data set of the pump fluence dependence}

In Fig.~\ref{fig:FigS8}a, we provide a second data set of the fluence dependence of the pump-induced terahertz electric field ($\Delta E$) at 7 K. Though the oscillations show slight differences compared to those in Fig.~\ref{fig:Fig2}b in the main text, the qualitative behavior is the same, specifically the softening of the mode energies (Fig.~\ref{fig:FigS8}b) and the initial linear rise in the amplitude of the modes followed by a decrease towards the critical fluence, as well as a crossing of the amplitudes of the two modes (Fig.~\ref{fig:FigS8}c).

\subsubsection{\normalsize B. Fits of the pump-probe response in the time domain}

As discussed in the main text, we fit the pump-probe response in the time domain using two damped sine waves. First, we show that a single damped sinusoid is not sufficient to describe the data. Figure~\ref{fig:FigS9} shows the pump-probe response at low and high fluences for both data sets with fits to a single damped sinusoid (red dashed lines) and two damped sinusoids (blue dashed lines). There is poor agreement between the data and the fits to a single oscillation, demonstrating the presence of more than one oscillation frequency. Instead, we can see that the sum of two damped oscillations is able to capture the salient features of the data. We remark that, while the fit matches the oscillations extremely well at initial pump-probe delay times, at later times there are slight deviations. These may be due to slight variations in the sample temperature caused by heat diffusion (see Fig.~\ref{fig:FigS7}). The fits were performed by taking into account the time resolution of the experiment, noting that the observed dynamics are a convolution of the oscillation model with the pump and probe pulse profiles (see Chapter 9 in Ref. \cite{prasankumar2016optical}).

We further note that the dynamics of the damped oscillations can be accurately tracked due to the lack of any relaxation background in the temporal trace. The latter is typically expected from the excitation and subsequent relaxation of charge carriers that are photodoped above the optical gap ($E_G$ $\sim$ 200 meV) \cite{gasparov2000infrared,randi2016phase}. Its absence signifies that, within the time resolution of our experiment ($\sim\,$100 fs), the excited charge carriers localize and assume a polaronic character, similar to what is observed in other correlated insulators governed by strong electron-boson coupling \cite{okamoto2011photoinduced}.

\subsubsection{\normalsize C. Fourier transform analysis of the pump-probe response}

\begin{figure}[htb!]
\includegraphics[width=\columnwidth]{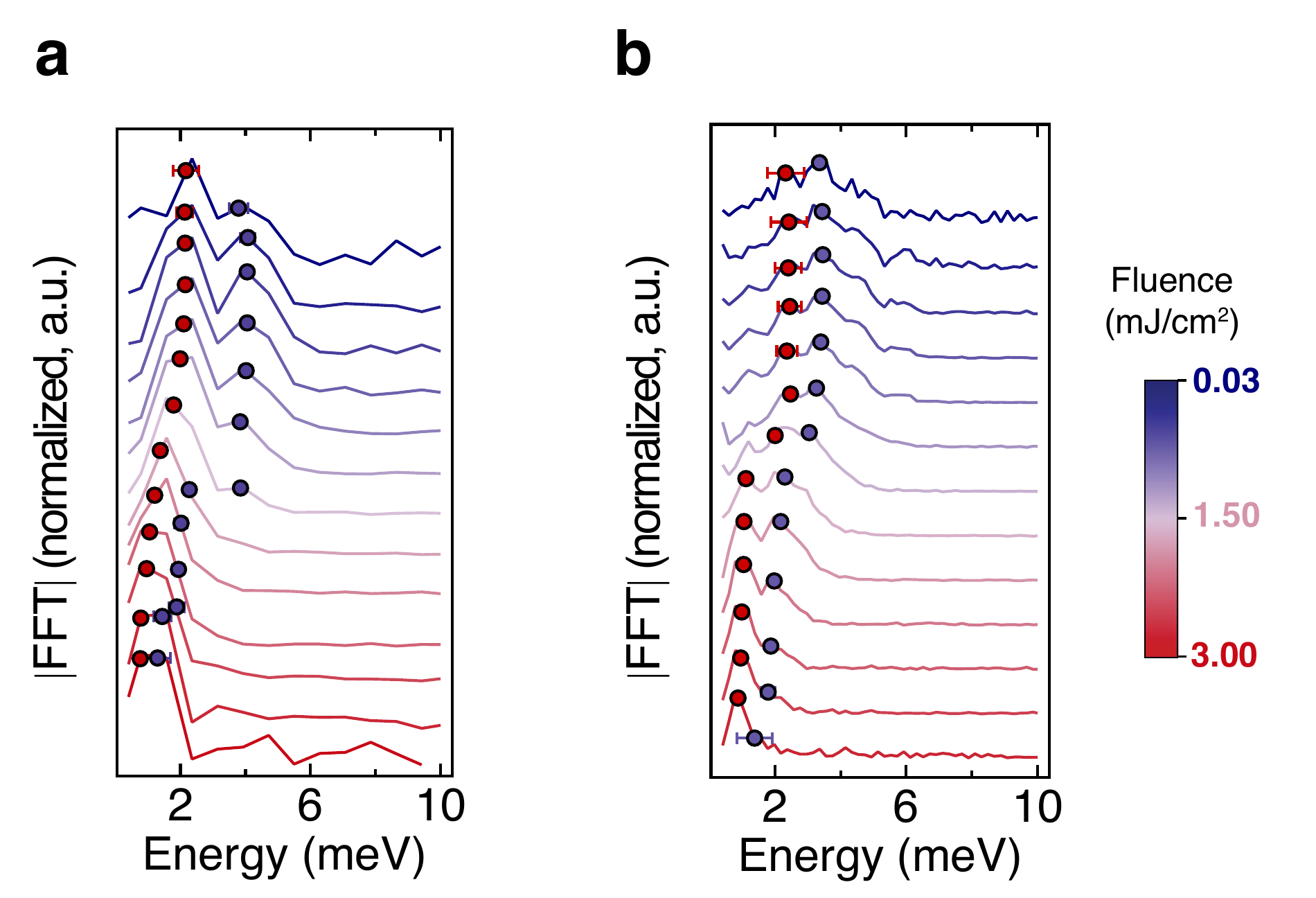}
\caption{\textbf{Fourier transform of the temporal traces.} \textbf{a}, Fourier transform analysis of each trace in Fig.~\ref{fig:Fig2}b in the main text. It is difficult to distinguish the two modes due to their close energies and large broadening. The red and violet dots represent the energies of the two collective modes obtained from the fits in Fig.~2c. The error bars indicate the 95\% confidence interval for these two fit parameters. The two methods for determining the energies of the modes agree well within the resolution of the Fourier transform and the error bars of the fits. \textbf{b}, Fourier transform analysis of each trace in Fig.~\ref{fig:FigS8}a along with the mode energies from the fits in Fig.~\ref{fig:FigS8}b. In this data set, it is easier to see two distinct peaks at higher fluences. As seen from both the fits in Fig.~\ref{fig:FigS9} and the Fourier transforms here, together these two data sets provide strong evidence for the presence of two coherent collective modes at all fluences.}
\label{fig:FigS10}
\end{figure}

A Fourier transform analysis confirms the results of the fits in the time domain. Figure~\ref{fig:FigS10} shows the Fourier transform of the two data sets with the fits to the mode energies (red and violet dots) superposed on the curves. It can be observed that the two methods for determining the energies of the modes agree well within the resolution of the Fourier transform and the error bars of the fits. This provides further evidence for the presence of two modes at all fluences.

\begin{figure*}[htb!]
\includegraphics[width=0.8\textwidth]{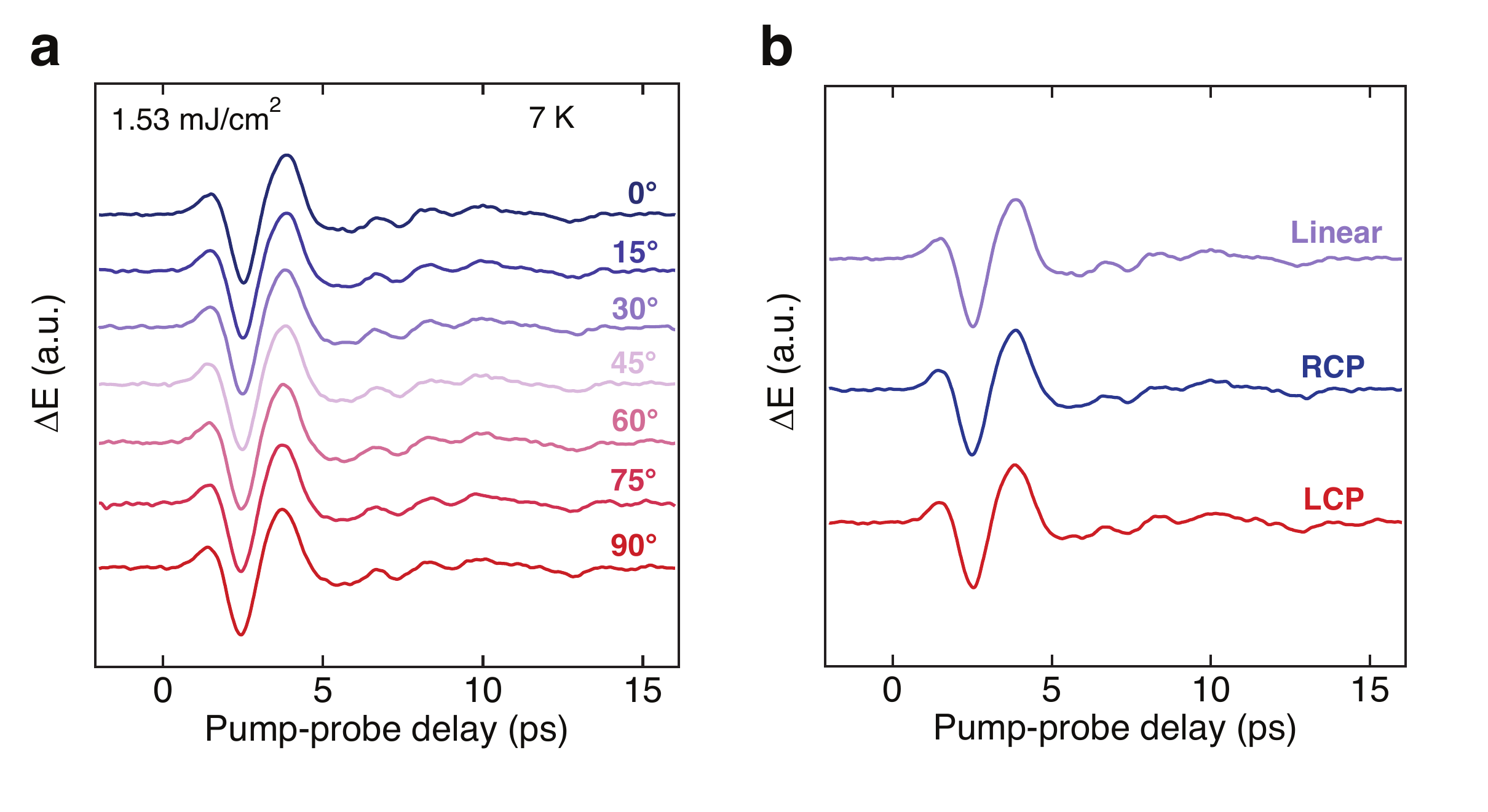}
\caption{\textbf{Pump polarization dependence.} \textbf{a}, Pump-probe response for different angles of the pump polarization. There is no change in $\Delta E$, indicating that the modes are totally symmetric. \textbf{b}, Response to a circularly polarized pump. The oscillations are identical when the pump beam is right and left circularly polarized (RCP and LCP) and linearly polarized as in \textbf{a}.}
\label{fig:FigS11}
\end{figure*}

\begin{figure}[t!]
\includegraphics[width=0.9\columnwidth]{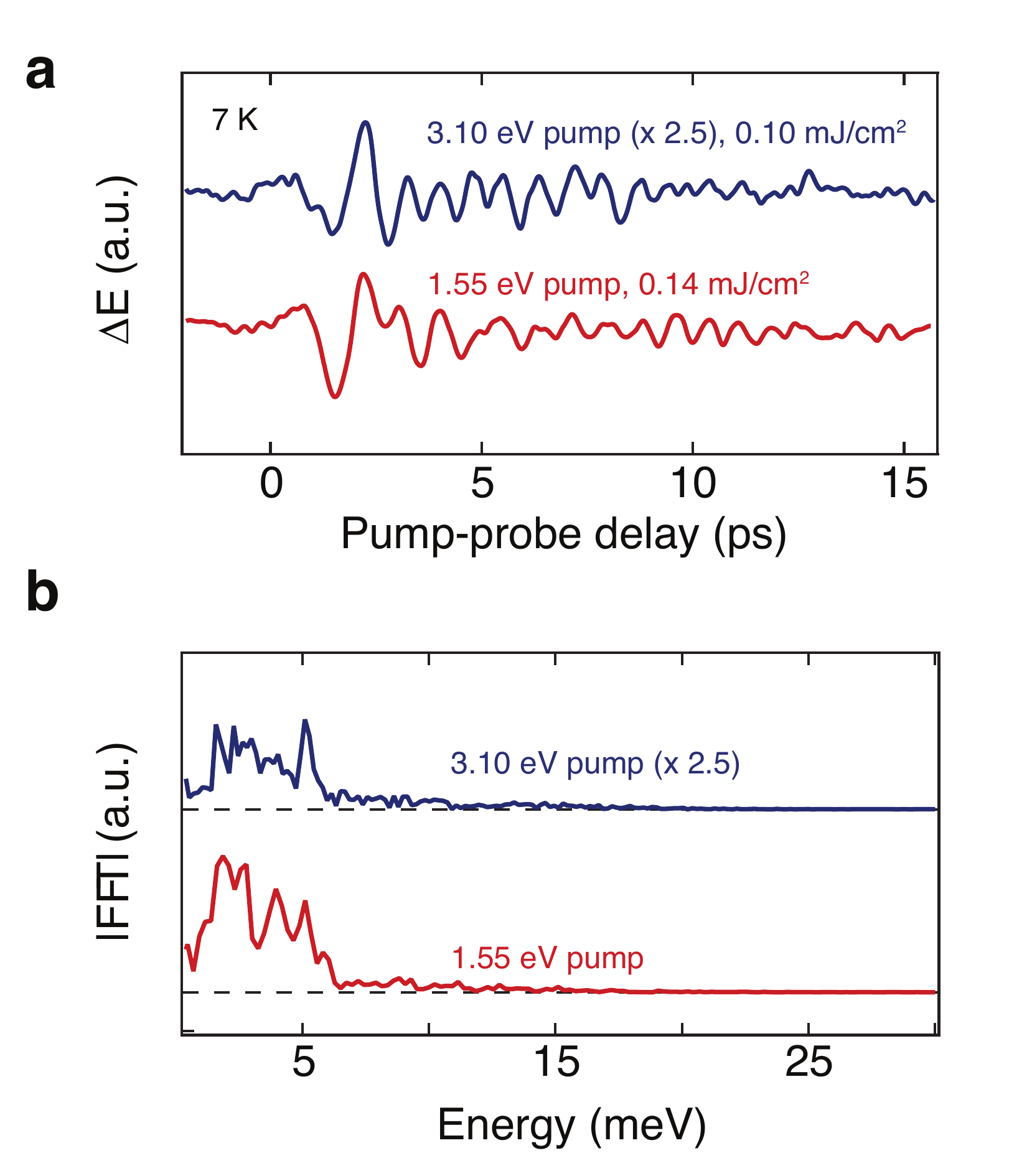}
\caption{\textbf{Response to 3.10 eV photoexcitation.} \textbf{a}, Pump-probe response when the pump photon energy is 3.10 eV (blue curve). The oscillations are very similar to those excited by the 1.55 eV pump (red curve). \textbf{b}, Fourier transform of both curves in \textbf{a}. There are no features at higher energies where a previous study observed totally-symmetric optical phonon modes excited by 3.10 eV light \cite{borroni2017coherent}. Our terahertz probe is therefore only sensitive to the low-energy electronic collective modes reported here.}
\label{fig:FigS12}
\end{figure}

\subsubsection{\normalsize D. Pump polarization dependence}

In order to assign the symmetry of the modes, we perform a pump polarization dependence (Fig.~\ref{fig:FigS11}a). The oscillations remain unchanged as the pump polarization is varied from parallel to perpendicular to the probe polarization. This isotropic response of the pump-probe signal indicates that the observed modes are totally symmetric. The same dependence was seen at all pump fluences. We also investigated the response to a circularly polarized pump (Fig.~\ref{fig:FigS11}b) and found that the same oscillations as in Fig.~\ref{fig:FigS11}a are present and do not change when the pump helicity is varied.

\subsubsection{\normalsize E. Pump-probe response with 3.10 eV excitation}

We also repeat the pump-probe experiments with a pump photon energy of 3.10 eV, using a BBO crystal to frequency double the 1.55 eV light from the laser. For this pump photon energy, which is close to the charge-transfer transition, we observe very similar oscillations to those excited by the 1.55 eV pump pulse (Fig.~\ref{fig:FigS12}a). Despite our $\sim\,$100 fs time resolution, the Fourier transform of the oscillations at this photon energy (Fig.~\ref{fig:FigS12}b) shows no signature of the totally-symmetric optical phonon modes in the range of 13.9 to 25.9 meV that have been observed in a previous study using a pump excitation of 3.10 eV \cite{borroni2017coherent}. This demonstrates that the spectral region of our terahertz probe is solely sensitive to the newly-discovered low-energy electronic modes.

\subsection{\normalsize Supplementary Note 6: Group theory aspects of the Verwey transition}

In this section, we perform a group theory analysis of the Verwey transition in magnetite in order to construct the simplest time-dependent Ginzburg-Landau (GL) model that is compatible with the symmetries of the system (see Methods). We remark that the analysis presented here is complementary to that of a previous study \cite{senn2013verwey}. In a phase transition from a high symmetry ($G$) to a low symmetry ($G_0$) space group, it is crucial to identify the irreducible representations (IRs) that lead to the symmetry breaking $G \rightarrow G_0$. This defines an inverse Landau problem \cite{ascher1977,hatch2001}. In magnetite, the space group above $T_V$ is $Fd\bar 3 m$ ($O^7_h$, 227), with cubic crystal symmetry and point group  $O_h$. Below $T_V$, the space group becomes $Cc$ ($C^4_{s}$, 9), with monoclinic crystal symmetry and point group $C_{s}$ \cite{yamauchi2009ferroelectricity,senn2012charge}. Therefore, we seek the IRs that lead to the symmetry breaking $ Fd\bar 3 m \rightarrow Cc $. We solve this problem with the aid of the software packages {\small GET$\_$IRREPS} \cite{aroyo2006,aroyo2006b,aroyo2011} and {\small ISOTROPY} \cite{stokes2007}. The transformation $\mathcal T$ relating the basis vectors of both phases is \cite{blasco2011}
\begin{equation}
\mathcal  T = \begin{pmatrix} 
1 & -1 & 0   \\
1 & 1& 0 \\
0 & 0 & 2 
\label{eq:transformation}
\end{pmatrix}. 
\end{equation}

\begin{figure*}[t!]
\includegraphics[width=6cm]{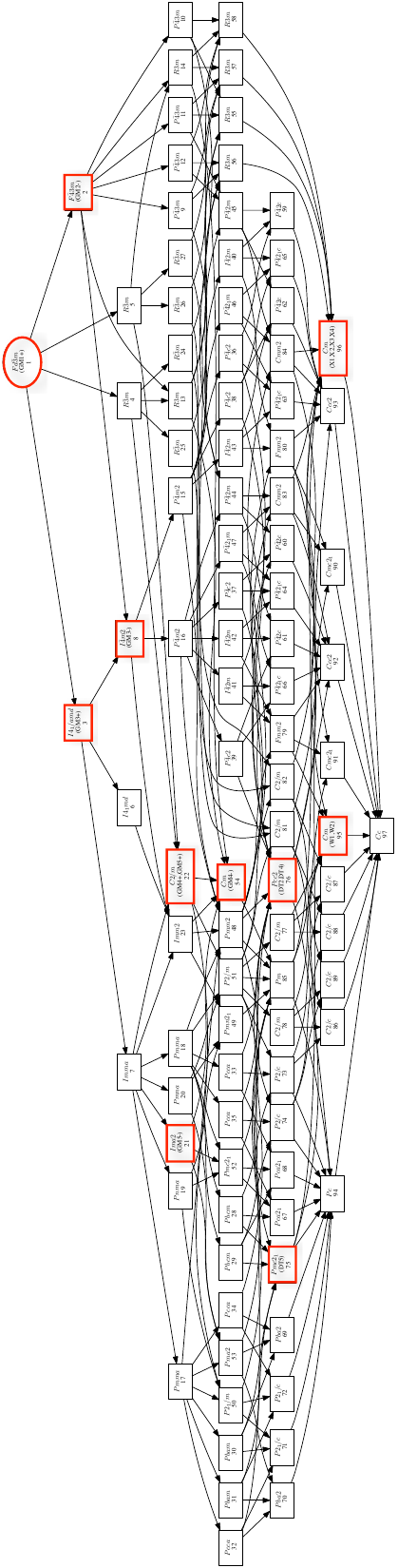}
\caption{\textbf{Group theory analysis of the Verwey transition.} Graph of isotropy subgroups obtained with the software suite {\small GET$\_$IRREPS} for the symmetry breaking $Fd\bar 3 m\rightarrow Cc$. Red boxes highlight the subgroups corresponding to IRs of the high-symmetry group.}
\label{fig:FigS13}
\end{figure*}

\noindent The isotropy subgroups are listed in Table~\ref{tab:isotropy}, along with the IRs and corresponding wave vectors. The main result is that none of the IRs give the low-temperature subgroup $Cc$. Therefore, we need at least two IRs to couple and condense in order to drive the transition. In principle, there are multiple choices of order parameters (OPs) that can drive the symmetry breaking $ Fd\bar 3 m \rightarrow Cc $, as seen in the graph of isotropy subgroups shown in Fig.~\ref{fig:FigS13}. From group theory arguments alone is not possible to determine which are the relevant IRs as all possible paths are allowed. However, experimental observations \cite{wright2002} and a previous group theory analysis \cite{piekarz2006mechanism} have identified that the OPs $X_3$ and $\Delta_5$ play a determinant role in the Verwey transition. Indeed, we verify that the intersection $Pmc2_1 \cap Cm$, corresponding to coupling the OPs  $X_3$ and $\Delta_5$ in a particular direction in representation space, generates the $Cc$ space group symmetry. Additionally, phonon modes with the symmetries $\Gamma$, $\Delta$, $X$, and $W$ have been identified to participate in the transition \cite{senn2012charge}, and all these IRs appear in Table~\ref{tab:isotropy}, in agreement with the experimental observations. 

Based on the result that a coupling between the IRs $X_3$ and $\Delta_5$ allows for the symmetry breaking $Fd\bar 3 m \rightarrow Cc$, some insight can be gained by studying the space group representation at the wave vectors $X$ and $\Delta$. The star of the \textit{k}-vector $X$ (obtained by applying the point group operations of $O_h$ to $X$) in the Brillouin zone has three arms: $(1,0,0), (0,1,0),$ and $(0,0,1)$ in units of $2\pi/a$, where $a$ is the lattice constant in the high temperature cubic unit cell. Since these \textit{k}-vectors are related by symmetry operations, the corresponding states are equivalent \cite{lee2011}. The group of the wave vector has 16 symmetry operations (not listed here) and four two-dimensional IRs $X_i$ for $i=1,2,3,4$ \cite{elcoro2017}. Among these four IRs, we assume that only $X_3$ participates in the transition with OP direction $(\eta_1,\eta_2,-\eta_2,-\eta_1,\eta_3,-\eta_3)$ in representation space as obtained with {\small ISOTROPY} (the dimension of representation space is defined by the number of arms in the star of $k$ times the dimension of the IR of the little group \cite{elcoro2017}). The arbitrary real constants $\eta_i$ with $i=1,2,3$ generate a three-dimensional subspace. From the direction of the OP in representation space, the wave vectors $\boldsymbol q^X_1 =(1,0,0), \boldsymbol q^X_2 = (0,1,0),$ and $\boldsymbol q^X_3 = (0,0,1)$ could be involved in the distortions causing the phase transition.

\begin{table}
\begin{tabular}{|c|c|c|}
\hline
 \; IRs   \;      &  Isotropy subgroup                  & $k$-vectors  ($2\pi/a$)    \\ \hline
 $ \Gamma^+_1$ &  $Fd\bar 3m$ (227)    & \multirow{8}{*}{ $ (0,0,0) $}  \\ \cline{1-2}
 $\Gamma^+_3$ &  $I4_1/amd$ (141) &   \\ \cline{1-2}
 $\Gamma^+_4$ &  $C2/m$ (12)       &   \\ \cline{1-2}
 $\Gamma^+_5$ &  $C2/m$ (12)       &   \\ \cline{1-2}
 $ \Gamma^-_2$ &  $F \bar 4 3 m$ (216)    &    \\ \cline{1-2}
 $\Gamma^-_3$ &  $I\bar 4m2$ (119) &    \\ \cline{1-2}
 $\Gamma^-_4$ &  $Cm$ (8)       &  \\ \cline{1-2}
 $\Gamma^-_5$ &  $Ima2$ (46)       &   \\ \hline
 $\Delta_2$         &  $Pcc2$ (27)        & \multirow{3}{*}{ (0,1/2,0)(1/2,0,0) \textbf{(0,0,1/2)}}  \\ \cline{1-2}
 $\Delta_4$         &  $Pcc2$ (27)        &   \\ \cline{1-2}
 $\Delta_5$         &  $Pmc2_1$ (26)        &  \\ \hline
 $X_1$                &  $Cm$ (8)       & \multirow{4}{*}{  \textbf{(0,1,0)} \textbf{(1,0,0)} \textbf{(0,0,1)}}  \\ \cline{1-2}
 $X_2$                &  $Cm$ (8)       &   \\ \cline{1-2}
 $X_3$                &  $Cm$ (8)       &  \\ \cline{1-2}
 $X_4$                &  $Cm$ (8)        & \\ \hline
 $W_1$                &  $Cm$ (8)        &\multirow{2}{*}{  $(1/2,1,0)\textbf{(1,0,1/2)}(0,1/2,1)$}\\ \cline{1-2}  
 $W_2$                &  $Cm$ (8)        &  \\ \hline  
\end{tabular}
\caption{\textbf{Irreducible representations that participate in the Verwey transition.} List of IRs, isotropy subgroups, and $k$-vectors between the parent group $Fd\bar 3 m$ and $Cc$. The $k$-vectors in bold are the subset of wavevectors of the irreducible representation involved in the symmetry-breaking process.}
\label{tab:isotropy}
\end{table}

On the other hand, the star of the \textit{k}-vector $\Delta=(0,2u,0)$ with $u=1/4$ has six arms: $(\pm1/2,0,0), (0,\pm1/2,0),$ and $(0,0,\pm1/2)$. The group of the wave vector has 48 symmetry operations and five allowed IRs $\Delta_i$ for $i=1, \cdots, 5$, four of which are one-dimensional and one is two-dimensional ($\Delta_5$). The relevant IR for the transition is $\Delta_5$ with OP direction $(0,0,0,0,0,0,0,0,\eta_4,\eta_5,\eta_5,-\eta_4)$ as obtained with {\small ISOTROPY}. Therefore, only the wave vectors $\boldsymbol q^\Delta_{4/5} = (0,0, \pm 1/2)$ are relevant for the transition. 

Using these results, we now construct an invariant polynomial under the space group operations of the high-symmetry group that couples the IRs $X_3$ and $\Delta_5$ \cite{stokes2007, hatch2003}, consistent with previous experimental observations \cite{wright2002,nazarenko2006,senn2012charge,de2013speed,kukreja2018}.  
The allowed terms in the polynomial are, to second-order, $F^{(2)} =|\boldsymbol X|^2+|\boldsymbol \Delta|^2$, where $\boldsymbol X = (X_1, X_2, X_3)$ and $\boldsymbol \Delta = (\Delta_1, \Delta_2)$.
The only allowed third-order term is $F^{(3)} =  X_1  X_2  X_3$, while the fourth-order terms are
\begin{align*}
F^{(4)} & =|\boldsymbol X|^4+|\boldsymbol \Delta|^4 + |\boldsymbol X|^2 |\boldsymbol \Delta|^2\\
 &+(4 X_1^4 + X_2^4 + 6 X_2^2 X_3^2 + X_3^4)\\ &+ (X_2^2 + X_3^2) (4 X_1^2 + X_2^2 + X_3^2)
 + X_1^2 (\Delta_1^2 + \Delta_2^2)\\ &+ (X_2^2 - X_3^2) \Delta_1 \Delta_2 +
(X_2^2 + X_3^2) \Delta_2^2 + (\Delta_1^4 +\Delta_2^4).
\end{align*}
\noindent This polynomial, which is the basis of a formal GL potential for the transition, involves five OPs related to the wave vectors that could, in principle, be involved in the transition. However, our experimental measurements are not momentum-resolved so the polynomial derived here is not directly relevant. Therefore, we construct a minimal GL potential based on this formal polynomial to describe the transition (see Methods).

\clearpage
\bibliography{Magnetite}

\end{document}